\documentclass[prd,superscriptaddress,showpacs,twocolumn,floatfix]{revtex4}
\input epsf
\usepackage{graphics}

\begin{document}
\title{The Dynamical Evolution of Black Hole-Neutron Star
  Binaries in General Relativity: Simulations of Tidal Disruption} 
\date{\today}
\author{Joshua A. Faber}
\altaffiliation{National Science Foundation (NSF) Astronomy and
  Astrophysics Postdoctoral Fellow} 
\email{jfaber@uiuc.edu}
\affiliation{Department of Physics, University of Illinois at
  Urbana-Champaign, Urbana, IL 61801} 
\author{Thomas W. Baumgarte}
\altaffiliation{J.S. Guggenheim Memorial Foundation Fellow}
\altaffiliation{Also at Department of Physics, University of Illinois at
  Urbana-Champaign, Urbana, IL 61801} 
\affiliation{Department of Physics and Astronomy, Bowdoin College,
  Brunswick, ME 04011} 
\author{Stuart~L.~Shapiro}
\altaffiliation{Also at Department of Astronomy and NCSA, University of
  Illinois at Urbana-Champaign, Urbana, IL 61801} 
\affiliation{Department of Physics, University of Illinois at
  Urbana-Champaign, Urbana, IL 61801} 
\author{Keisuke Taniguchi}
\affiliation{Department of Physics, University of Illinois at
  Urbana-Champaign, Urbana, IL 61801} 
\author{Frederic A. Rasio}
\affiliation{Department of Physics and Astronomy, Northwestern
  University, Evanston, IL, 60208} 

\date{\today} 

\begin{abstract}
We calculate the first dynamical evolutions of
merging black hole-neutron star binaries that construct the combined black
hole-neutron star spacetime in
a general relativistic framework.  We treat the metric in the
conformal flatness approximation, and assume that the black hole mass
is sufficiently large compared to that of the neutron star so that
the black hole remains fixed in space.  Using a spheroidal spectral
methods solver, we solve the resulting field
equations for a neutron star orbiting a Schwarzschild black hole.
The matter is evolved using a relativistic, Lagrangian, smoothed
particle hydrodynamics (SPH) treatment. We take as our initial data
recent quasiequilibrium models for synchronized neutron star polytropes
generated as solutions of the conformal thin-sandwich (CTS) decomposition of
the Einstein field equations.  We are able to construct from these
models relaxed SPH configurations whose profiles show good agreement
with CTS solutions.
Our adiabatic evolution calculations for neutron stars with
low-compactness show that
mass transfer, when it begins while the neutron star orbit is still
outside the innermost stable circular orbit, is more unstable than is
typically predicted by analytical formalisms.  This dynamical mass
loss is found to be the driving force in determining the subsequent
evolution of the binary orbit and the neutron star, which typically
disrupts completely within a few orbital periods.  The majority of the
mass transferred onto the black hole is accreted promptly; a
significant fraction ($\sim 30\%$) of the mass is shed outward as well, some of
which will become gravitationally unbound and ejected completely from
the system.  The remaining portion forms an accretion disk around the
black hole, and could provide the energy source for short-duration
gamma ray bursts. 
\end{abstract}
\maketitle

\section{Introduction}

The infall of compact objects into black holes (BHs) is of
considerable interest in many branches of astrophysics.  In
particular, many of the arguments that can be made about coalescing
neutron star-neutron star (NSNS) binaries also apply to coalescing
black hole-neutron star (BHNS) binaries.  Both are strong candidates
for the central engines of short-duration gamma ray bursts (GRBs), since
the merger timescale following tidal disruption 
is comparable to the GRB duration and the 
gravitational binding energies provide the 
characteristic energy scales inferred
by observers \cite{JERF,RossGRB}.  It is possible that any
ejected matter may contribute significantly to the r-process elemental
abundance of the universe \cite{FRT,Ross,RSW}.  Additionally, they are
expected to be among the most important sources of gravitational waves
(GWs) that can be detected by both terrestrial laser interferometers
such as LIGO \cite{LIGO}, VIRGO \cite{VIRGO}, GEO \cite{GEO}, and TAMA
\cite{TAMA}, as well as the proposed space-based interferometer LISA
\cite{LISA1}.

The key difference between the sources that can be observed with LIGO
(and comparable detectors) and LISA is the characteristic frequency of
the GW emission: LISA's characteristic frequency range falls within
$10^{-4}-10^{-1}~{\rm Hz}$, whereas LIGO operates between $10-500~{\rm
Hz}$.  Because of this, LIGO is most sensitive to the mergers of
stellar-mass BHs, whereas LISA will observe more massive merging
systems that involve either intermediate mass BHs (IMBHs), $M_{\rm
BH}=10^2-10^4 M_{\odot}$, or supermassive BHs (SMBHs), $M_{\rm BH} >
10^5 M_{\odot}$.  The formation history leading to these encounters is
likely to involve completely different processes.  

Compact binaries with stellar-mass BHs are likely to be formed through
typical stellar binary evolution, at rates that depend on 
parameters such as the binary mass ratio distribution, 
common-envelope efficiency, and the physics of supernova kicks, all of
which remain somewhat uncertain
(see \cite{KNST,BKB,VT} and references therein for a thorough review).
The mass distribution of BHs in such systems is poorly constrained, as
none have been observed to date, but may vary widely, spanning a range
$2M_\odot<M_{\rm BH}<25 M_{\odot}$ \cite{BSR}.
For sufficiently tight binaries, merger will occur within a Hubble
time.  In these cases, the dissipative effects of gravitational
radiation will cause the orbit to circularize as the binary separation
shrinks, so that the eccentricity of the orbit is expected to be
almost zero by the time the binary enters the LIGO band.  Whether or
not the compact object is tidally disrupted by its BH companion, as
well as where this would occur in the latter case with respect to 
the Innermost Stable Circular Orbit (ISCO), depends
on both the compaction of the compact object and the mass ratio (see
Section \ref{sec:physical}).

This simple picture does not apply to compact objects orbiting BHs
with considerably higher mass.  Both IMBHs and SMBHs are expected to
reside within stellar clusters, whose dynamics will be determined by
both stellar-BH and stellar-stellar gravitational encounters
(scattering).  
Some stars will typically be scattered, either strongly or weakly, into the
``loss cone'', i.e., the volume of phase space encompassing orbits
with sufficiently small periastrons that the star will be tidally
disrupted before being kicked into another orbit by future
encounters (see \cite{Shap85} for a review of the original derivations, and
\cite{Sig,Merritt} for more recent work).  As a result, most objects
that enter the loss cone do so at very high eccentricity, with
periastron distances of $5-50$ Schwarzschild radii
\cite{Frei1,HA}.  In many cases, these systems will approach the
BH with eccentricities $e\gtrsim 0.1$ \cite{BaC}.

GW detections from coalescence with higher
mass BHs may yield very little information about the physics of NS
matter, for the case of a NS falling into an IMBH, or any
compact object (BH, NS, or white dwarf) falling into an SMBH with
$M\gtrsim 10^6 M_{\odot}$. These objects should plunge through the
ISCO of the BH intact, since the
tidal-disruption radius lies {\it within} the ISCO, and will likely be
swallowed whole by the BH.  For the opposite
case, applicable to white dwarfs (WD) falling into IMBHs (and NSs into
stellar-mass BHs), tidal disruption will occur outside the ISCO, a
process we describe in detail in Sec.~\ref{sec:physical}.

For the vast majority of its lifetime, a stellar-mass compact object
binary will inspiral very slowly, such that it can be described by a
point-mass, post-Newtonian (PN) treatment.  PN formalisms for the
adiabatic inspiral epoch are now completely determined up to 3.5PN
order \cite{BDEI1}, and include lowest-order spin-orbit and
spin-spin terms \cite{Kidder,Will}.  Once finite-size and tidal
effects become important at close separation, it becomes necessary to
solve the fully nonlinear Einstein field equations.
Quasi-equilibrium binary configurations in circular orbits 
have been calculated in GR
for NSNS \cite{BCSST,BCSST2,BGM2,GGTMB,SU,Duez2,TG}, BHNS
(see \cite{BSS}, hereafter BSS; \cite{Keisuke}, hereafter TBFS, and
references therein), and BHBH binaries (\cite{Cook,Baum,GGB1,Yo};
for a thorough review of the topic and references, see
\cite{CookLR,BS}).  Details of 
the transition from slow inspiral to rapid plunge, and deviations from
the point-mass energy versus frequency relation found in
quasi-equilibrium sequences, may yield important information about the
physical parameters of the NS equation of state (EOS; see, e.g.,
\cite{BCSST,OT,FGRT}).  It has been suggested \cite{Hughes}
that a combination of $~10-50$ broadband and narrowband
observations of NSNS mergers might be able to constrain the NS radius
to within a few percent.  We will show below that BHNS mergers may be
just as interesting, but it is likely that the interpretation of
physical features in the GW signal will be significantly more
complicated, since differences between stable and unstable modes of
mass transfer may lead to radically different scenarios.

Eventually, for those systems in which the tidal limit is reached
before the ISCO, mass transfer onto the BH will begin.  This process
is fundamentally dynamic in nature, and can only be modeled accurately
by relativistic, three-dimensional hydrodynamic calculations.
Attempts have been made to model these systems analytically, but as we
will show below, the conclusions rely on a number of unphysical
assumptions.  The earliest work describing mass transfer in detail for
compact object binary mergers \cite{CE} assumed that mass transfer in
NSNS binaries would conserve both mass and orbital angular momentum,
and that both NSs would remain on a quasi-circular orbit in corotation
during the
evolution; a similar set of assumptions was used to describe BHNS
binaries as a possible source of gamma-ray bursts \cite{PZ}.  A more
complex treatment developed in \cite{DLK} drops the assumption of
circularity, since it is not seen to hold in numerical calculations
(e.g., \cite{Ross}).  Still, their model for the evolution of BHNS
systems undergoing mass transfer depends on a number of ad hoc
assumptions that need to be tested by dynamical calculations in order
to be proven valid.

Beyond uncertainties about the form of the late-inspiral GW signal
produced by a BHNS merger, there remains the question of the event
rate, which remains uncertain given the complete lack of detection of
such systems to date.  Still, it is possible to estimate the likely
merger rate using population synthesis models, which can be calibrated
against the observed galactic NSNS binary population and supernova
rates.  Recent estimates predict an advanced LIGO annual detection
rate of anywhere from a few mergers \cite{VT} up to potentially
several hundred \cite{BKB}.

Should a BHNS binary merger be observed, it might reveal a great deal
about the physics of matter at nuclear densities.  In particular, the
onset of mass transfer would yield a clear indication about the NS
radius, and, as we will explain in detail below, the stability of the
mass transfer would yield important information as to the nuclear EOS.
Whereas for NSNS binaries the characteristic frequencies of GW
emission during the merger and formation of a remnant (either a
hypermassive NS or BH) will typically occur at frequencies outside the
peak sensitivity of even an advanced LIGO detector, the same is not
true for BHNS binaries.  Since the frequency at the onset of
instability scales roughly inversely with the total binary mass, we
expect stellar-mass BHNS mergers to occur at characteristic
frequencies at which LIGO will be most sensitive, $\sim 100-500~{\rm
Hz}$.  If the GW signal from a merger was observed to be coincident
with a short-duration gamma-ray burst, we could potentially determine
their distance, luminosity, and characteristic beaming
angle \cite{KM}.  A detailed theoretical understanding of these
systems is now more urgent than ever, in light of the recent
localizations of short GRB afterglows
\cite{SGRB,Gehrels,Covino,Fox,Berger,050813}, 
the first ever for
these systems (many long-duration GRBs have been localized, but are
believed to be the result of collapsing stars, not merging compact
binaries).

Unfortunately, the current state-of-the-art for hydrodynamic
calculations of BHNS inspiral and merger is far behind that for NSNS
mergers.  Calculations of the latter have been performed using a
variety of Newtonian, PN, and relativistic gravitational formalisms
(see, \cite{RS,BS} for thorough reviews, and \cite{FGR}, hereafter
FGR, for a more recent summary).  Many calculations have now been
performed in either the conformal flatness (CF) approximation to
general relativity (GR) \cite{FGR,Oech}, or in full GR
\cite{SU1,SU2,STU1,STU2}.  These GR calculations now include
sophisticated treatments of the NS EOS and physically appropriate
initial spins (\cite{STU2}; NSs are expected to be nearly irrotational
in the inertial frame, since the viscous timescale is much longer than
the inspiral timescale, see \cite{BC,Koch} and Sec.~\ref{sec:physical}
below).

The key difficulty that must be overcome to perform simulations of
relativistic BHNS mergers is the same one that arises 
in the study of BHBH binaries; the presence of a spacetime singularity
inside the black hole. To avoid encountering the singularity in a
numerical simulation, the BH interior is
excised from the computational grid in most current applications.
This is justified by the fact that no information can 
propagate from the BH interior to the exterior, so the
exterior can be evolved independently of the interior.  While 
progress has been reported, especially very 
recently \cite{P05}, black hole evolution
calculations have been plagued by numerical instabilities.  In some
ways, BHNS mergers are even more difficult to evolve consistently,
since both the singular behavior of the BH as well as the hydrodynamic
nature of the NS must be confronted.  Whereas the BHBH problem involves a
pure vacuum solution of the Einstein field equations, the NS must
always be evolved in such a way that the relativistic fluid is treated
properly.

As a result of these difficulties, all hydrodynamic calculations
performed to date of stellar-mass BHNS mergers have used Newtonian or
quasi-Newtonian gravitational treatments
\cite{LK0,LK1,LK2,LK3,LK4,JERF,RSW}.  Needless to say, binaries
containing a BH can be evolved accurately only by using relativistic
hydrodynamics in a relativistic spacetime.  We emphasize here that
this applies both to the tidal field created by the BH, as well as the
self-gravity of the NS.  Previous Newtonian calculations have in some
cases \cite{LK2,LK4} used an approximate black hole potential,
suggested in \cite{PW}, that creates an ISCO at
$6M_{\rm BH}$, but no single static potential can generate the full
set of relativistic forces experienced by matter in the strong-field
regime.  
Calculations employing a fixed background BH metric have typically been
performed for stars undergoing a tidal interaction with a massive
BH, rather than a stellar-mass BH, with relativistic dynamical terms
but a Newtonian treatment of the self-gravity. The secondary, in fact,
is often assumed to be a white dwarf or main-sequence star.
These models include
SPH treatments without self-gravity \cite{Laguna}, and both PPM
\cite{FroDien} and spectral method \cite{Marck} treatments
with Newtonian self-gravity.
More recently, SPH techniques have been
devised that evolve the NS matter in the background metric of a
stellar-mass BH,
using Newtonian-order correction to model the NS self-gravity, for
both SPH \cite{PSBH,PSBH2} and characteristic gravity \cite{Bishop}.  
This approach is appropriate for describing
main-sequence stars or white dwarfs.  However, since the tidal
disruption is a result of a competition between the black hole tidal
force and stellar self-gravity, this approach is not sufficient to
describe BHNS binaries accurately.  Modeling tidal disruption in
BHNS binaries requires a relativistic treatment of both the 
black hole and the neutron star.

Here, we will make use of the CF approximation to GR, introduced by Isenberg
\cite{Isen} and Wilson and collaborators \cite{WMM}.  The CF
approximation amounts to assuming that the spatial metric remains
conformally flat, so that the gravitational fields can be found by
solving the constraint equations of GR, decomposed in the conformal
thin-sandwich (CTS) decomposition \cite{Yor99}, alone.  The CTS
formalism has been used in numerous applications to
construct initial data describing both NSNS and BHNS binaries in
quasiequilibrium
\cite{BCSST,BCSST2,Duez2,SU,GGTMB,BGM2,TG,BSS,Keisuke}.
For these initial data the choice of a conformal background metric is
completely consistent with Einstein's initial value (constraint) field
equations, although different
choices may describe the astrophysical situation at hand more or less
accurately.  The situation is different for dynamical simulations in
the CF approximation (e.g.~\cite{WMM,Oech,FGR,Dimm1,
VPCG}), since the assumption that the spatial metric {\em
remains} conformally flat is no longer strictly consistent with Einstein's
field evolution equations.  For many applications, however, CF provides an
excellent approximation.  For spherically symmetric configurations,
as an example, the CF approach is exact, and for many other
applications the error has been shown to be in the order of at most a
few percent (see, e.g., \cite{CooST96}). It is particularly useful for
exploring dynamical behavior, e.g., collapse or tidal break-up, which
occurs on dynamical timescales and is unaltered by secular effects
like gravitational radiation-reaction.

In this paper we present the dynamical extension of BSS, who
calculated the first relativistic,
quasiequilibrium BHNS sequences as solutions of the CTS
decomposition of the Einstein field equations.  Modifying
their code to treat the metric for the Schwarzschild BH in isotropic (CF)
coordinates, rather than the Kerr-Schild coordinates reported in BSS,
we take their corotating quasiequilibrium configurations as initial
data.  As in BSS, we assume an extreme mass ratio, $M_{\rm BH}\gg
M_{\rm NS}$, which allows us to hold the BH position fixed and
restrict the computational grid to a neighborhood of the NS,
thereby avoiding complications arising in the BH interior.
We also assume a polytropic equation of state for
the neutron star, as well as synchronous rotation.  The resulting
dynamical calculations are the first of their kind to solve the
CTS field equations for the spacetime around the NS self-consistently by
treating both the NS 
and BH relativistically.  They allow us to study details
of the dynamical mass-transfer process, particularly its
stability.  The CF approximation holds a stable equilibrium
configuration constructed in the CTS formalism in strict dynamical
equilibrium.  Our calculation is a prototype of more detailed
general relativistic 
calculations we hope to provide in the future that will involve
irrotational NS models with more realistic EOSs and compactions,
arbitrary mass ratios, and a fully self-consistent treatment of the
spatial metric.

In the CF approximation, gravitational radiation reaction
must be added in by hand in order to drive the system toward merger.
While it is the secular energy losses to gravitational radiation that
initially drive the binary system toward the point of tidal
disruption, they play a much reduced role in the dynamics thereafter.
Indeed, while secular forces determine the path the binary takes prior
to merger, the merger itself is a fundamentally dynamical process, as
we discuss in great detail below. 

Our work is organized as follows.  In Sec.~\ref{sec:physical} we
discuss the important physical scales that define our problem, and
present a detailed treatment of the traditional picture for determining
the stability of mass transfer.  We then discuss the limitations of
this model, and explain why it may not be applicable for BHNS mergers.
In Sec.~\ref{sec:numerical} we describe our numerical methods,
including the details of both our implementation of the CF field
equations as well as our use of smoothed particle hydrodynamics (SPH)
techniques to evolve the fluid configuration.  In
Sec.~\ref{sec:equil} we compare our relaxed initial data to previous
quasiequilibrium models, 
and find that we can construct configurations that satisfy
the field equations to high accuracy while reproducing previous
results.  In Sec.~\ref{sec:dynamical} we present our simulations of
merging binaries for different models of the NS polytropic EOS.  Finally, in
Sec.~\ref{sec:discussion} we discuss our results in the context of GW
astrophysics, and describe our plans for further calculations.

\section{Physical Overview}\label{sec:physical}

The evolution of binaries containing NSs is a fully
relativistic problem, since lowest-order PN approximations break down
in the strong gravitational fields present
during late stages of the merger.  However, 
we can use information from Newtonian and
quasi-Newtonian calculations to estimate the various timescales and
physical regimes we expect to encounter.  Thus, we first classify the
relevant physical scales we expect to encounter in our study of BHNS
binary evolution, and then generalize the standard model for stable,
binary mass transfer to relativistic stars.  
In doing so, we will explain why this model, variants of
which have been used previously to describe the evolution of compact
binaries, is unlikely to apply to BHNS mergers.

Our simple mass-transfer model does have physical relevance, as it can apply
to the case of a WD inspiraling in a nearly circular fashion
toward an IMBH in a globular cluster. Such a star will begin transferring
mass long before reaching the ISCO.  However,
since WDs are typically kicked into highly eccentric orbits prior to
interactions with the BH, the orbit may not have time to
circularize fully before the onset of mass transfer.  
In such cases, the binary evolution will
be more complicated than the scenario we consider here; it has been studied
before by several groups (see \cite{Sig,RBB}, and references therein).

\subsection{Units, Timescales and Characteristic Lengths}

The four
most important timescales characterizing the problem at hand are the NS
dynamical timescale $t_D$, the viscous timescale $t_v$, the orbital
timescale $T$, and the GW radiation-reaction timescale $t_{\rm
GW}$.  Throughout this paper, we set
$G=c=1$.  The BH and NS masses can be written in terms of the initial
mass ratio $q$ as $M_{\rm BH}=q^{-1}M_{\rm NS}$ or equivalently
$M_{\rm NS}=qM_{\rm BH}$, and the NS radius $R_{\rm NS}={\cal
C}^{-1}M_{\rm NS}=q{\cal C}^{-1}M_{\rm BH}$, where ${\cal C}\equiv
M_{\rm NS}/R_{\rm NS}$ is the compactness parameter.

The dynamical timescale of the NS is given by
\begin{equation}
t_D\equiv \sqrt{\frac{2R_{\rm NS}^3}{M_{\rm NS}}}=2^{1/2}{\cal
  C}^{-1.5}M_{\rm NS}=
2^{1/2}q{\cal C}^{-1.5}M_{\rm BH}.
\end{equation}
We wish to compare this with the orbital and radiation-reaction
timescales at the radius where Roche-lobe overflow will begin.  To
estimate this radius, we will use the approximate form proposed in
\cite{PacRoche},
\begin{equation}
R_r = 0.46a\left(\frac{q}{1+q}\right)^{1/3},\label{eq:rroche}
\end{equation}
which gives the (volume-averaged) Roche-lobe radius as a function of
the mass ratio and binary separation $a$, for a binary treated as a
pair of point masses in the corotating frame.  This definition differs
from the original definition of the Roche lobe, which was defined for
incompressible matter (point masses are in effect infinitely
compressible), but the physical scalings are the same; the coefficient
becomes $0.41$ for incompressible matter 
instead of $0.46$. The Roche-lobe radius is equal
to the NS radius at a separation
\begin{eqnarray}
a_R&=&2.17q^{-1/3}(1+q)^{1/3}{\cal C}^{-1}M_{\rm
  NS}\nonumber\\
&=&2.17q^{2/3}(1+q)^{1/3}{\cal C}^{-1}M_{\rm BH}, 
\end{eqnarray}
at which point the (Keplerian) orbital period is
\begin{eqnarray}
T\equiv2\pi\sqrt{\frac{a_R^3}{M_T}}&=&20.1(1+q)^{1/2}{\cal
  C}^{-1.5}M_{\rm NS}\nonumber\\
&=&20.1q(1+q)^{1/2}{\cal
  C}^{-1.5}M_{\rm BH}\sim 14 t_D,
\end{eqnarray}
where $M_T\equiv M_{\rm NS}+M_{\rm BH}$ is the total binary mass, and 
the last relation holds for $q\ll 1$, or equivalently, $M_T\simeq
M_{\rm BH}$.  We note that in this
limit, the orbital period is a fixed multiple of the NS dynamical
timescale, regardless of the properties of the NS.

For a point-mass binary on a circular orbit, lowest order radiation
reaction predicts that the binary will inspiral, losing energy and
angular momentum, on a characteristic timescale $t_{\rm GW}$ given by
\begin{equation}
t_{\rm
    GW}=\frac{a}{\dot{a}}=\frac{E}{\dot{E}}=
    \frac{J}{2\dot{J}}=\frac{5}{64}\frac{a^4}{M_{\rm 
    NS}M_{\rm BH}M_T}=4\tau,\label{eq:tgw}
\end{equation}
where $\tau$ is the coalescence time, i.e., the remaining time until
the point-mass binary would reach $a=0$. At the Roche-lobe separation,
assuming $q\ll 1$, we have
\begin{eqnarray}
t_{\rm GW}&=&1.73 q^{2/3}{\cal C}^{-4}M_{\rm NS}\nonumber\\
&=&1.73 q^{5/3}{\cal
  C}^{-4}M_{\rm BH}= 
1.23 q^{2/3}{\cal C}^{-5/2}t_D. 
\end{eqnarray}
In general, the radiation-reaction timescale will be at least an order
of magnitude longer than the dynamical timescale for any binary which
begins mass transfer outside the ISCO radius.  
The orbital period $T$, however, may become similar to the coalescence
time $t_{\rm GW}$, indicating that the infall becomes quite rapid.

\begin{figure}[ht]
\centering \leavevmode \epsfxsize=\columnwidth \epsfbox{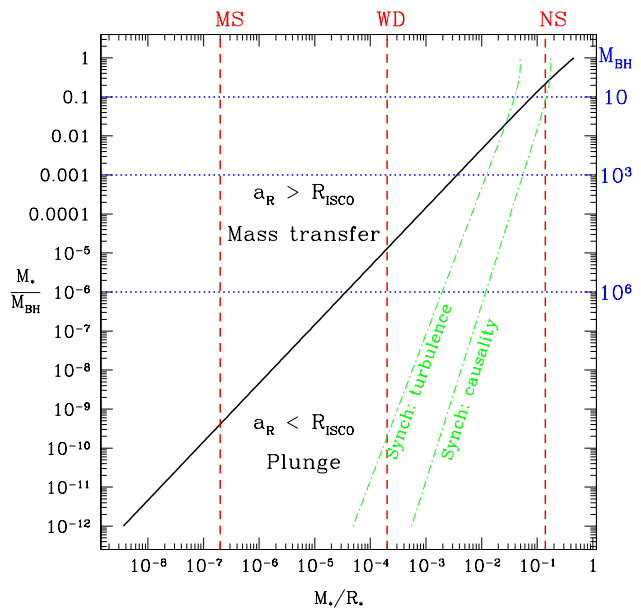}
\caption{The critical mass
ratio $q = M_*/M_{\rm BH}$ as a function of the secondary's 
compaction ${\cal C} =
M_*/R_*$, for which mass transfer begins at the ISCO, taken here as $a
= 6M_{\rm BH}$ (solid line). For systems above the curve, mass
transfer begins while the orbit is stable; for those
below, the secondary may plunge into the BH before being tidally
disrupted.  Dashed vertical curves show characteristic compactions of
a WD or NS; dotted horizontal curves show typical mass ratios for a $1
M_{\odot}$ compact object orbiting a $10 M_{\odot}$ stellar-mass BH, a
$10^3 M_{\odot}$ IMBH, or a $10^6 M_{\odot}$ SMBH.  Dot-dashed curves
show where $\beta$, the ratio of the light crossing time the
viscous timescale of the NS, equals unity, and where $\alpha_{\rm Vis}$,
the nondimensional turbulent viscosity assumes its
largest reasonable value for the turbulent
viscosity [See Eqs.~(\protect\ref{eq:betav}) and
  (\protect\ref{eq:alphav})].  
Only configurations to the left of these curves can
synchronize prior to merger.}
\label{fig:qc}
\end{figure}

In Fig.~\ref{fig:qc}, we show the regions in parameter space for which the
critical separation for Roche-lobe overflow lies within or outside of
the ISCO, for a wide variety of compact object-BH binaries.  Dashed
vertical lines correspond to the approximate compactness of either a WD or a
NS, whereas dotted horizontal lines show the approximate mass ratios
to be expected for a $10M_{\odot}$ stellar-mass BH, an IMBH, or a
SMBH.  Systems above the critical curve reach the tidal limit before
the ISCO, and are likely to transfer mass onto the BH.  For those
sufficiently below the curve, we expect that the compact object will
pass through the ISCO intact, and plunge onto the BH relatively
intact.

We note however, that this simple picture may very well be altered by
a number of more complicated effects.  Recently, Miller \cite{Miller}
has argued that even if systems are expected to reach the
mass-shedding limit prior to crossing the ISCO, in many cases they will
have already begun to plunge.  Indeed, since the binary energy as a
function of separation flattens out significantly near the ISCO,
$t_{GW}$, which is already nearly of order $T$, 
will systematically underestimate the infall velocity
(a similar argument was used in \cite{FGRT} to argue that the GW energy
spectrum produced in NSNS mergers declines dramatically near the
ISCO).  Because of this, the ISCO may {\it systematically
underestimate} the binary separation at which prompt merger becomes
inevitable.

On the other hand, describing the ``plunge'' of an extended object
like a NS may provide a misleading picture of the dynamical merger
in cases where the tidal disruption occurs inside the ISCO but outside
the horizon.
While it is certain that some matter, likely a significant fraction of
the NS mass, will plunge inward directly onto the BH, this may liberate a
great deal of angular momentum into the outer parts of the NS
\cite{Ross05}.  As a result, some fraction of the mass may survive the
plunge, at least initially, in the form of a ``mini-NS'', which will
escape outside the ISCO on an elliptical orbit.  Needless to say, only
dynamical calculations will clarify the role played by these competing
effects.

The final timescale we must consider is the viscous damping timescale
$t_{\rm Vis}$, 
which we expect to play a crucial role in determining the fate of the
binary once mass transfer commences.  In the limit that the viscous
timescale is extremely short (high viscosity), we expect two important
phenomena to occur.  First, tidal dissipation can synchronize the
binary so that the secondary is corotating upon the onset of mass
transfer.  From \cite{BC}, we note that a binary will be synchronized
by the time mass transfer begins only if 
\begin{eqnarray}
\beta\equiv\frac{R_{\rm NS}}{t_{\rm Vis}}&\gtrsim
&60(1+q)^{5/3}q^{-2/3}{\cal C}^3, \label{eq:betav}\\
t_{\rm Vis} &\lesssim&\frac{1}{60}(1+q)^{-5/3}q^{2/3}{\cal C}^{-4}M_{\rm
NS}\nonumber\\
&\lesssim&\frac{1}{60}(1+q)^{-5/3}q^{5/3}{\cal C}^{-4}M_{\rm BH},
\end{eqnarray}
where $\beta$ is the ratio of the light crossing time of the secondary to its
viscous timescale, as defined by \cite{BC}.  If we follow \cite{Duez}
and assume that turbulent viscosity is the primary damping mechanism, we can
define $\alpha_{\rm Vis}$, the turbulent viscosity parameter, so that
\begin{equation}
\alpha_{\rm Vis}\equiv\frac{t_D}{t_{\rm Vis}}\label{eq:alphav}.
\end{equation}
We see that synchronization will occur if 
\begin{equation}
\alpha_{\rm Vis}\gtrsim 60(1+q)^{5/3}q^{-2/3}{\cal C}^{5/2}.
\end{equation}
On Fig.~\ref{fig:qc}, we show curves marking the critical mass
ratio-compactness dependence for $\beta=1$, which we define as
the ``causal limit'', as well as for $\alpha_{\rm Vis}=0.1$, which is 
the maximum plausible value for turbulent viscosity in physical systems
of interest \cite{BH}.  Configurations to the left of the curve can synchronize
before merger; this includes essentially all mergers where the
secondary is either an MS star or a WD.  NS mergers, on the other
hand, will be irrotational in general, especially when the primary is
a BH, since the required viscosity to synchronize the NS increases as
the primary mass increases \cite{BC,Koch}.
Viscosity should also play a role after mass transfer starts, as we
will discuss in detail below.  

\subsection{The stability of mass transfer}\label{sec:stability}

Once the secondary fills its Roche lobe, it will begin to transfer
mass onto its companion.  Such a process can be either stable or
unstable, depending on its response to mass loss.  If the volume of
the Roche lobe shrinks faster than (or expands slower than) the
stellar radius, the process is unstable, and the star will typically
be disrupted violently.  On the other hand, if a small amount of mass
loss causes the star to shrink back within the Roche lobe, it is
possible for the mass loss to temporarily cease, or at the very least
settle down to a much smaller equilibrium level, whose value can be
determined based on the assumptions made about conservation of mass
and angular momentum, as we discuss below.

We first note that models of stable mass transfer typically assume
that the binary orbit remains quasicircular, which in turn is only
possible if the viscous timescale is short relative to the orbital and
GW timescales.  Maintaining a circular orbit requires that the orbital
energy evolve according to a fixed relation in terms of the orbital
angular momentum and the mass of the secondary, but there is no reason
to assume that such a relation should hold {\it a priori}.  Indeed, it
is viscous dissipation that drives the orbit toward circularity, by
converting excess orbital energy into other forms.  Furthermore, when
the viscous timescale is long, the mass-transfer rate can grow 
extremely rapidly, since the inner Lagrange point
travels into the secondary at roughly $v_{in}R_{NS}/a_R$, unbinding
progressively denser material from the NS. If this leads to an
unstable runaway, it is the mass loss that drives the orbital
evolution, and we expect to find the development of an orbital
eccentricity.  This violates the typical assumptions made in
conservative mass-transfer models, which assume that mass loss is
steady, and slow enough that the orbit can remain circular as mass is
lost.
 
The early attempt to follow the mass transfer process in detail for
NSNS binaries was provided by \cite{CE}, who modeled
the heavier NS as a point mass, and the lighter secondary using an EOS
that yields a nearly flat mass-radius relation down to $M_{\rm NS}\sim
0.3M_{\odot}$, below which the NS begins to expand rapidly with
further decreasing mass.
Rather than assume conservative mass transfer,
they parameterized the possible loss of both mass and angular momentum
from the system, finding that the former has very little effect on
their results.  In their model, mass transfer leads to a widening of
the binary orbit, under the condition that the NS radius must equal
the radius of its Roche lobe.  As mentioned above, this will only hold
for systems in which $t_{\rm Vis} \ll t_{\rm GW}$.  Over time, the mass loss
rate and GW luminosity decrease rapidly from their large initial
values at closest approach
(as does the rate of neutrino production as the NS matter
decompresses during the transfer), until eventually the low-mass NS
begins to expand rapidly \cite{CST1} and unstable mass transfer begins.
 
Many of these ideas were revisited for a discussion of BHNS binaries
in \cite{PZ}, in light of the optical identification
of GRB counterparts at cosmological distances.  Assuming a Newtonian
$n=1.5$ polytropic EOS and fully conservative mass transfer, they find
that the initial mass-transfer rate between a $1.4 M_{\odot}$ NS and a
BH with mass $M_{\rm BH}=3-5 M_{\odot}$ will occur at a rate of $\sim
100 M_{\odot} {\rm s}^{-1}$ for approximately $1~{\rm ms}$ before
decaying away according to the approximate power-law relation
$\dot{M}_{\rm NS}\propto t^{-14/11}$, which corresponds to $M_{\rm
NS}(t)\propto t^{-3/11}$.  As in \cite{CE}, they assume that the
process will terminate when the NS reaches a critical minimum mass and
begins to expand unstably.
 
The most recent treatment of BHNS coalescence makes a completely
different set of assumptions about the dynamics during mass transfer.
Based on the Newtonian BHNS numerical calculations of Rosswog, Speith,
and Wynn \cite{RSW}, Davies, Levan, and King \cite{DLK} assume that
the rapid timescale for mass transfer will violate the assumption of
circular orbits, which underlies the typical conservative,
quasiequilibrium mass-transfer
formulation.  Instead, they make the following assumptions:
\begin{enumerate}
\item Mass transfer occurs during a timescale corresponding to half an
  orbit.
\item During this time period, the NS, treated as a uniform density
  sphere, will lose mass from a shell whose depth is a distance
equivalent to the
  infall rate from the beginning of the mass-transfer rate multiplied
  by half an orbital period.
\item Half of the angular momentum lost to the transferred mass will
  return to the NS, placing it on an eccentric orbit that will
  typically not lead to overflow during the next periastron passage.
\end{enumerate}
This model does reproduce well the extremely high mass loss rates
initially seen during Newtonian numerical calculations of BHNS mergers
\cite{LK1,JERF,RSW}, but the assumptions adopted are
somewhat ad hoc. In particular, transferring angular momentum back to
the NS without adjusting its mass causes a discontinuous evolution of
the binary orbit.  In some cases, the NS will find itself on an orbit
whose {\it periastron} is outside the mass-shedding limit, leading to
a period of stable evolution until GW dissipation forces the orbit to
decay inward again back to the onset of mass transfer.  In contrast,
we find below that mass transfer can be quenched temporarily, 
but from this point on the
NS follows an elliptical trajectory that will take it back within the
mass-shedding limit prior to the next periastron passage.

In Appendix~\ref{appendix:masstransfer}, we derive a semi-analytic
formulation for
{\it conservative} mass transfer onto a BH, modeling secondaries 
either by a Newtonian or a  relativistic polytrope.  We
recover the scaling relations found in \cite{CE,PZ}, and
generalize them for arbitrary polytropic indices.  Although these
relations are unlikely to hold for merging BHNS systems, as shown in
Fig.~\ref{fig:qc}, the
Newtonian results can be applied to merging WDs as well as main
sequence stars undergoing mass transfer.  We note that there are semi-analytic
formalisms for describing {\it non-conservative} mass transfer as well
(see, e.g. \cite{PJH} for a formalism involving mass transfer from a
main sequence star onto a companion), and that these have been useful
in describing WDBH mergers \cite{FWHD}, but that the actual NS tidal
disruption process is sufficiently dynamic that essentially all
analytic treatments break down.

The theory of accretion disk dynamics presents several interesting
connections to that of merging binaries, since questions about the
stability of mass transfer appear as well (see, e.g., \cite{DF} and
references therein).  One key difference between the models is the
typical radial angular momentum distribution; parameterizing the
tangential velocity profile as $v_t(r)\propto r^{\alpha}$,
irrotational NS have a nearly flat velocity profile,
$\alpha\sim 0$, and corotating NS a flat angular velocity profile,
$\alpha=1$, both larger than the
Keplerian value $\alpha_K=-0.5$.  Moreover, NS differ greatly from
disks because of their infall velocity when they pass through the ISCO, 
and their strong self-gravity.  While angular
momentum distributions with larger values of $\alpha$ help to
stabilize mass transfer in
disks, stronger self-gravity destabilizes mass transfer \cite{DF}.  Thus, it is hard
to generalize across the classes, although we note that irrotational
NS should, if anything, be more prone to unstable mass transfer, as
the NS loses more angular momentum per unit mass lost from its inner edge.

\section{Numerical techniques}\label{sec:numerical}

To compute the dynamical evolution of a BHNS binary,  
we fix the position of the BH
and assume that the surrounding spacetime metric takes the
form appropriate to a nonspinning Schwarzschild BH.  The
approximation of a fixed BH position
is correct in the limit that $M_{\rm BH}\gg M_{\rm NS}$.
Here, we will study binaries with mass ratios $q\equiv M_{\rm
NS}/M_{\rm BH}=0.1$, which is presumably within the range of values
for which the approximation of an extreme mass ratio is valid.

To calculate gravitational forces and evolve the fluid configuration,
we will work within the CF formalism, which we explain in more detail
in Section \ref{sec:cf} below.  We assume that the spatial metric is
remains conformally flat, so that it can be
written in the form
\begin{equation}\label{eq:cfmetric}
ds^2=(-\alpha^2+\beta_k\beta^k)dt^2+2\beta_idx^idt+\psi^4\delta_{ij}dx^idx^j,
\end{equation}
where $\alpha$ and $\beta^i$ are the lapse function and shift vector,
respectively.  Under this assumption we only need to solve the
$3+1$ constraint equations for $\psi$, $\alpha$ and $\beta^i$ to
determine the metric.

Our initial configuration places the NS in a {\em
corotating} initial configuration.  Irrotational configurations, which
are more realistic astrophysically, will be treated in a later publication.
We model the NSs as relativistic polytropes, and assume adiabatic
evolution, which we describe in detail in Sec.~\ref{sec:hydro}.

The code we use both to relax and evolve BHNS binaries is similar to
that introduced in FGR \cite{FGR} 
for evolving NSNS binaries.  We solve the five
linked non-linear field equations of the CF formalism,
Eqs.~(\ref{eq:shift}), (\ref{eq:psi}), and (\ref{eq:alphapsi}) below, using
the LORENE libraries, publicly available at {\tt
http://lorene.obspm.fr}.  These Poisson-like equations are solved
using spectral methods, decomposing the fields and their sources in a
set of radially distinct domains into radial and angular expansions.
Dynamical evolution is treated through 
SPH discretization.  Many aspects of the code were
discussed in detail in FGR, so we concentrate instead on the changes
and new features introduced to evolve BHNS binaries.

Roughly speaking, we have made three significant changes to the code
to admit the presence of a BH in the binary.  First, the asymptotic
Schwarzschild 
BH contribution to the spacetime metric is held fixed, allowing us to
solve the field equations describing the self-gravity of the NS in a
fully consistent way.  Second, as discussed below, we solve
Poisson-like elliptic equations for $\psi$ and $(\alpha\psi)$, as in BSS and
elsewhere, rather than for $\nu\equiv\ln \alpha$ and $\beta\equiv\ln
(\alpha\psi^2)$, as in FGR and related treatments (TBFS denotes the latter
quantity ``$\sigma$'').  Third, we restrict the spatial domain of our
spectral methods field solver to a finite radius centered on the NS,
as was done in BSS and TBFS, which allows us to avoid problems
 near the BH. Indeed, our computational domain is chosen so as not to
 overlap the event horizon at any time.   As a
result, we do not make use of the asymptotic boundary conditions
typically used by LORENE-based codes, which can be extended to spatial
infinity through the proper coordinate transformations \cite{BGM}.
The use of a restricted spatial domain has been introduced before, in
the context of domains with ingoing and outgoing GWs \cite{NB}, but
with a set of BC's that are not appropriate to the (elliptic) problem at hand.
Instead, as we describe below, we have introduced a multipole
expansion BC, used here and in TBFS, which should be more accurate
than the lowest-order power-law falloff conditions traditionally used
in grid-based calculations.  Below, we first summarize the relevant
equations that comprise relativistic hydrodynamics
(Sec.~\ref{sec:hydro}) and the CF formalism (Sec.~\ref{sec:cf}),
introduce the ``split'' equations which factor out the BH
contributions to the spacetime in Sec.~\ref{sec:bhnssplit}, describe
our new approach for introducing a multipole BC in
Sec.~\ref{sec:multipole}, and finally describe how this affects the
evaluation of various quantities in the SPH evolution equations
Sec.~\ref{sec:sph}.

\subsection{Relativistic Hydrodynamics} \label{sec:hydro}

We assume that the matter can be
described as a perfect fluid so that the stress-energy tensor takes 
the form
\begin{equation}
T^{\mu\nu}=\rho_0(1+\varepsilon+\frac{P}{\rho_0})u^\mu u^\nu +
Pg^{\mu\nu},\label{eq:perfflu}
\end{equation}
where $\rho_0$, $\varepsilon$, $P$, and $u^\mu$ denote the rest mass
density, specific internal energy, pressure, and 4-velocity,
respectively.  We will describe the NS by a relativistic polytropic
EOS that evolves adiabatically with index $\Gamma$.  Hence,
the pressure obeys the relation
\begin{equation}
P=(\Gamma-1)\rho_0\varepsilon,\label{eq:polytrope}
\end{equation}
and initially satisfies
\begin{equation}
P=\kappa\rho_0^\Gamma,\label{eq:polyinit}
\end{equation}
where $\kappa$ is a constant.  As discussed in BSS, we can
scale away dimensional units by setting $\kappa=1$ (see their Sec.~IIIc).

The Lagrangian continuity equation (FGR,\cite{Oech}) can be written as
\begin{eqnarray}
\frac{d\rho_*}{dt}+\rho_* \partial_i v^i=0\label{eq:cont}
\end{eqnarray}
where we define the conserved density 
\begin{equation}
\rho_*\equiv \alpha u^0\psi^6\rho_0=\gamma_n \psi^6\rho_0,
\end{equation}
and the coordinate velocity 
\begin{equation}
v^i=\frac{u^i}{u^0}=-\beta^i+\frac{u_i}{u^0\psi^4},\label{eq:vofutilde}
\end{equation}
and introduce the Lorentz factor for the fluid $\gamma_n\equiv\alpha
u^0$.  Lagrangian time derivatives are related to Eulerian partial
time derivatives through the familiar relation
$d/dt\equiv\partial/\partial t+v^i\partial_i$.
To determine the Lorentz factor, we solve the normalization
condition for the 4-velocity, 
\begin{equation}
{\gamma_n}^2=(\alpha u^0)^2=1+\frac{u_i
u_i}{\psi^4}=1+\frac{\tilde{u}_i\tilde{u}_i}{\psi^4}\left[1+\frac{\Gamma \kappa
\rho_*^{\Gamma-1}}{(\gamma_n \psi^6)^{\Gamma-1}}\right]^{-2},\label{eq:gamman}
\end{equation}
implicitly.

The Euler equation can be written
\begin{equation}
\frac{d\tilde{u}_i}{dt}=-\frac{\alpha\psi^6}{\rho_*}\partial_i P -
\alpha hu^0\partial_i \alpha+\tilde{u}_j\partial_i
\beta^j+\frac{2h\alpha (\gamma_n^2-1)}{\gamma_n \psi}\partial_i \psi,
\label{eq:dutildedt}
\end{equation}
where the specific momentum is defined by
\begin{equation}
\tilde{u}_i\equiv hu_i,
\end{equation}
and the specific enthalpy $h$ by
\begin{equation}
h\equiv 1+\Gamma \varepsilon.
\end{equation}

Finally, the energy equation takes the form
\begin{equation}
\frac{de_*}{dt}+e_*\partial_i v^i=0,\label{eq:destardt}
\end{equation}
where $e_*=\gamma_n\psi^6 (\rho_0 \varepsilon^{\Gamma-1})^{1/\Gamma}$.
For an adiabatic evolution without shock heating, the energy equation
is satisfied automatically by adopting Eq.~(\ref{eq:polyinit}).

To account for shocks, we included an
artificial viscosity prescription composed of both
linear and quadratic terms (the relativistic analogue of the form
introduced in \cite{HK}, similar to that found in \cite{Oech}).  We
found no evidence for significant shocks within the body of the NS,
as only the matter in the mass transfer stream directed toward the BH
showed signs of significant heating very near the BH.  Using the
value of $\kappa\equiv 
P/\rho_0^\Gamma\equiv(\Gamma-1)\varepsilon/\rho_0^{\Gamma-1}$ 
as a measure, a quantity that remains constant during
an adiabatic evolution, we found variation of no more than $5\%$
within the body of the NS.  This is hardly a surprise, as there is no
physical mechanism such as a collision to cause significant shocking
within the bulk of the NS.
Shock heating will be important for
understanding the evolution of the initially low-density accretion stream that 
falls toward the BH, especially near the event horizon.  In this
region, the heating can be substantial, but it seems not to introduce
significant feedback on the NS remnant.
Given these results, we replace the energy
equation, Eq.~(\ref{eq:destardt}), with its adiabatic 
solution, Eq.~(\ref{eq:polyinit}),
throughout the calculations described here.  In future calculations,
where shocks may be more important, we will restore the full
evolution of the energy equation with an artificial viscosity
prescription and allow for shocks everywhere. 
This will be especially important for irrotational NS calculations,
since the matter transferring through the inner
Lagrange point has significantly greater angular momentum than in the
irrotational case, and a great deal of it will likely forming a disk
rather than accreting promptly.

\subsection{The CF formalism}\label{sec:cf}

In the CF formalism \cite{Isen,WMM} we assume
that the spatial metric is not only conformally flat initially, but
that it remains conformally flat.  In particular, for the 3-metric 
we approximate
$\partial_t \tilde \gamma_{ij} = 0$ so that in rectangular coordinates
$\tilde \gamma_{ij} = \delta_{ij}$ at all times.  Strictly speaking, 
this is inconsistent with Einstein's evolution equations, but is 
often a very good approximation, particularly on dynamical timescales
when secular motion due to radiation-reaction is not important.  Under
this approximation the evolution equation for the spatial metric
yields a relation between the extrinsic curvature and the shift,
\begin{equation}
K_{ij}\equiv\frac{\psi^4}{2\alpha}\left[\delta_{il}\partial_j\beta^l+
\delta_{jl}\partial_i\beta^l-\frac{2}{3}\delta_{ij}\partial_l\beta^l\right].
\end{equation} 
Inserting this expression into the momentum constraint yields an equation
for the shift $\beta^i$
\begin{eqnarray}
\nabla^2 \beta^i+\frac{1}{3}\partial^i(\partial_j \beta^j)&=&
16\pi\alpha\psi^4(E+P)U^i\nonumber\\
&&\!\!\!\!\!\!\!\!\!\!\!\!\! +2\alpha\psi^4K^{ij}\nabla_j(\ln[\alpha/\psi^6])\equiv S_{\beta}^i,
\label{eq:shift}
\end{eqnarray}
where $\nabla^2$ is the flat space Laplacian.
The Hamiltonian constraint is an equation for the conformal factor $\psi$
\begin{equation}
\nabla^2\psi=-2\pi\psi^5E-\frac{1}{8}\psi^5K_{ij}K^{ij}\equiv S_{\psi}.
\label{eq:psi}
\end{equation}
To derive an equation for the lapse $\alpha$, the remaining
undetermined function in the 
metric, Eq.~(\ref{eq:cfmetric}), we choose maximal
slicing $K \equiv \gamma^{ij} K_{ij} = 0$ at all times, which implies
$\partial_t K=0$.  
This choice can be combined with the evolution equation for the
extrinsic curvature, which then yields
\begin{equation}
\nabla^2(\alpha\psi)=2\pi \alpha\psi^5(E+2S)+
\frac{7}{8}\alpha\psi^5K_{ij}K^{ij}\label{eq:alphapsi}\equiv S_{\alpha\psi}.
\end{equation}
In the above equations the matter sources $E$, $S$ and $U^i$ are projections
of the stress-energy tensor $T_{\mu\nu}$ and can be expressed as
\begin{eqnarray}
E&=&\rho_0 h \gamma_n^2-P,\\
S&=&3P+\frac{\gamma_n^2-1}{\gamma_n}(E+P),\\
U^i&=&\frac{\tilde{u}_i}{\gamma_n h \psi^4}.
\end{eqnarray}

In practice, it is easier to decompose the three coupled equations for
the shift, Eq.~(\ref{eq:shift}), into four decoupled Poisson equations.
To do so, we follow  \cite{ONS,SBS} and define
\begin{equation}\label{eq:bchi}
\beta^i=4B_i-\frac{1}{2}[\partial_i(\chi+B_kx^k)]=
\frac{7B_i-\partial_i\chi-x^k\partial_iB_k}{2}, 
\end{equation}
and solve the set
\begin{eqnarray}
\nabla^2 B_i&=& \frac{S_{\beta}^i}{4},\\
\nabla^2 \chi &=& -\frac{S_{\beta}^i x_i}{4}.
\end{eqnarray}

These Poisson-like equations, found in \cite{SBS} and elsewhere, are
exactly equivalent to those found in FGR for $\nu\equiv \ln\alpha$ and
$\beta\equiv \ln(\alpha\psi^2)$, and share the same asymptotic
fall-off behavior, but have radically different properties near the
horizon of the BH, where the lapse function goes to zero.  This causes
divergences in the values of $\nu$ and $\beta$, whereas $\alpha$ and
$\psi$ remain finite and easy to deal with in a numerical treatment.
Our chosen variables also exhibit a slightly different behavior when we
split them into additive pieces contributed largely by the NS and BH,
i.e., the contributions from the NS and BH to $\psi$ and
$\alpha\psi$ are additive, whereas the logarithmic dependence of the
``$\nu-\beta$'' set means that the two contributions are combined
multiplicatively.

As several different sets of notation have now been introduced into
the literature to define equivalent quantities in the CF formalism, we
present alternate notations used in a selection of other works in
Table~\ref{table:notation}.

\begin{table*}[th!]
\caption{A comparison of our notation for various relativistic
quantities to that found in a selection of 
previous works using the CF formalism:
\protect\cite{FGR,GGTMB,Oech,SBS,WMM}.  For those cases where no unique
terminology was defined, we give the simplest equivalent algebraic form.}
\begin{tabular}{l|cccccc}
Quantity & Here & FGR \protect\cite{FGR} & Gourgoulhon
\protect\cite{GGTMB} & Oechslin \protect\cite{Oech} & Wilson
\protect\cite{WMM}& Shibata \protect\cite{SBS}\\
\colrule\colrule
Lapse & $\alpha$ & $N$ & $N$ & $\alpha$ & $\alpha$ & $\alpha$ \\
Shift & $\beta_i$ & $-N_i$ & $-N_i$ & $\beta_i$ & $\beta_i$ & $\beta_i$ \\
Conformal Factor & $\psi$ & $\sqrt{A}$ & $\sqrt{A}$ & $\psi$ & $\phi$ & $\psi$ \\
Rest Density & $\rho_*$ & $\rho_*$ & $\Gamma_n A^3 \rho$ & $\rho_*$ &
$D\phi^6$ & $\rho_*$ \\
Lorentz Factor & $\gamma_n$ & $\gamma_n$ & $\Gamma_n$ & $\alpha u^0$ &
$W$ & $\alpha u^0$\\
Velocity & $v^i$ & $v^i$ & $NU^i+N^i$ & $v^i$ & $V^i$ & $v^i$ \\
Specific Momentum & $\tilde{u}_i$ & $\tilde{u}_i$ & $w_i$ & $\tilde{u}_i$
& $S_i/(D\phi^6)$ & $\tilde{u}_i$ \\
Enthalpy & $h$ & $h$ & $h$ & $w$ & $h$ & $1+\Gamma \epsilon$ \\
\colrule
\end{tabular}
\label{table:notation}
\end{table*}

\subsection{BHNS binaries}\label{sec:bhnssplit}

The CF approximation is exact for spherically symmetric
configurations, reproducing the TOV equation for fluid configurations
as well as the Schwarzschild solution for a stationary, non-spinning
black hole.  In isotropic coordinates, such a solution is given by
\begin{equation}
ds^2=-\left(\frac{1-M_{\rm BH}/2r}{1+M_{\rm BH}/2r}\right)^2
dt^2+\left(1+\frac{M_{\rm BH}}{2r}\right)^4\delta_{ij}dx^i dx^j.
\end{equation}
From this metric we identify the BH lapse and conformal factors as
\begin{eqnarray}
\alpha_{{\rm BH}}&=&\frac{1-M_{\rm BH}/2r}{1+M_{\rm BH}/2r},
\label{eq:alphabh}\\
\psi_{{\rm BH}}&=&1+\frac{M_{\rm BH}}{2r}.\label{eq:psibh}
\end{eqnarray}
The BH contribution to the shift $(\beta_i)_{{\rm BH}}$, vanishes in
isotropic coordinates (unlike in the Kerr-Schild coordinates used by
BSS).

To convert this line element to the more familiar Schwarzschild
(areal) coordinates, with
\begin{equation}
ds^2=-\left(1-\frac{2M_{\rm
    BH}}{\tilde{r}}\right)dt^2+\left(1-\frac{2M_{\rm
    BH}}{\tilde{r}}\right)^{-1}d\tilde{r}^2+ 
\tilde{r}^2d\Omega^2,
\end{equation}
one makes the coordinate transformation
\begin{eqnarray}
\tilde{r}&=&\left(1+\frac{M_{\rm BH}}{2r}\right)^2r\label{eq:rbarr},\\
r&=&\frac{1}{2}\left[\tilde{r}-M_{\rm
    BH}+\sqrt{\tilde{r}(\tilde{r}-2M_{\rm BH})}\right]. 
\end{eqnarray}
Note that this implies that the Schwarzschild radius and ISCO radius
take the values
\begin{eqnarray}
\tilde{r}=2~M_{\rm BH} &\leftrightarrow& r=0.5~M_{\rm BH},\\
\tilde{r}=6~M_{\rm BH} &\leftrightarrow& r=4.949~M_{\rm BH}.
\end{eqnarray}
At asymptotically large distances, $\tilde{r}=r+M_{\rm BH}$.

As discussed at length in \cite{GGTMB}, it is useful to ``split'' the
field equations when dealing with binaries, so that the terms on the
RHS of each equation are concentrated on either component of the
binary.  Here, the method is slightly different.  Since the BH
solution is an exact solution of the vacuum field equations, it can be
subtracted out of the full metric field equations to yield the CF
solution for the largely-NS contribution to the fields.  Defining
$N\equiv\alpha\psi$, and accordingly, $N_{\rm BH}\equiv \alpha_{\rm
BH}\psi_{\rm BH}=1-M_{\rm BH}/{2r}$, we split the fields such that
\begin{eqnarray}
\psi&=&\psi_{{\rm BH}}+\psi_{{\rm NS}},\\
\alpha&=&\alpha_{{\rm BH}}+\alpha_{{\rm NS}},\\
N\equiv\alpha\psi&=&N_{{\rm BH}}+N_{{\rm NS}}\\  
&\Rightarrow& N_{{\rm NS}}=
\alpha_{{\rm NS}}\psi_{{\rm NS}}+\alpha_{{\rm BH}}\psi_{{\rm
    NS}}+\alpha_{{\rm NS}}\psi_{{\rm BH}}.\nonumber 
\end{eqnarray}
The NS piece of the field equations, Eqs.~(\ref{eq:psi}) and
(\ref{eq:alphapsi}), can be expressed as
\begin{eqnarray}
\nabla^2\psi_{{\rm NS}}&=&-2\pi(\psi_{{\rm BH}}+\psi_{{\rm NS}})^5E\nonumber\\
&&-\frac{1}{8}(\psi_{{\rm BH}}+\psi_{{\rm NS}})^5
(K_{ij})_{{\rm NS}}(K^{ij})_{{\rm NS}},\\
\nabla^2N_{{\rm NS}}&=&2\pi (N_{{\rm BH}}+N_{{\rm NS}})(\psi_{{\rm
    BH}}+\psi_{{\rm NS}})^4(E+2S)\nonumber\\ 
&&\!\!\!\!\!\!\!\!\!\!\!\!\!\!\!\!\!\!\!\!\!\!\!\!
+\frac{7}{8}(N_{{\rm BH}}+N_{{\rm NS}})(\psi_{{\rm BH}}+\psi_{{\rm NS}})^4
(K_{ij})_{{\rm NS}}(K^{ij})_{{\rm NS}}.
\end{eqnarray}
The BH contributes to the shift vector, Eq.~(\ref{eq:shift})
only through the lapse function and conformal factor, since the black hole
contribution to the shift vanishes in isotropic coordinates.

\subsection{Multipole Boundary Conditions}\label{sec:multipole}

The LORENE-based field solver we use decomposes the angular dependence
of all scalar, vector, and tensor quantities into spherical harmonics
(the radial decomposition into Chebyshev polynomials is described in
detail in \cite{BGM}).  For configurations in which the outermost
boundary extends to spatial infinity, the outer boundary condition can
be set exactly to zero for any field which satisfies a power-law
falloff. For a BHNS binary, however, that is not an option, since we
encounter numerical difficulties when the computational domain
overlaps the BH singularity.  Instead, we must impose an approximate
BC for each field on the outermost (spherical) boundary, which lies at
a finite radius, as shown in Fig.~(\ref{fig:decomp}).  This outermost
boundary is chosen so that it never overlaps the BH event horizon.

Any Poisson-like equation $\nabla^2\Phi=\rho$ with compact support has
an exterior solution given by
\begin{eqnarray}
\Phi(\tilde{r},\theta,\phi)&=&\sum_{l=0}^\infty\sum_{m=-l}^l\left[\int_0^R
  \rho r^l Y^*_{lm}(\theta,\phi)d^3 \vec{r}\right]\nonumber\\&&\times Y_{lm}(\theta,\phi)
\tilde{r}^{-(l+1)}\nonumber\\
&\equiv& \sum_{l=0}^\infty\sum_{m=-l}^l
\rho_{lm}Y_{lm}(\theta,\phi)\tilde{r}^{-(l+1)},\label{eq:multexp} 
\end{eqnarray}
where we define the multipole moments of the source term $\rho_{lm}$
[see Eq.~(4.2) of \cite{Jackson}].
This solution established the boundary conditions for our outermost
computational domain, and can be matched to the interior solutions to
yield the field solution everywhere in space.  Here, the source terms
of the Poisson-like Eqs.~(\ref{eq:shift}), (\ref{eq:psi}),
and (\ref{eq:alphapsi}) are not compact, but instead satisfy rather steep
power-law falloffs, allowing us to use the same formalism while
introducing only small errors.  
Noting that the matter configurations are equatorially
symmetric, we only sum over multipoles with the same equatorial
symmetry as the particular field (i.e., $l+m$ even for $\psi$,
$\alpha$, $\beta^x$, $\beta^y$, and $\chi$; $l+m$ odd for $\beta^z$).
Rather than evaluate the real field source integrals against
the complex spherical harmonics $Y_{lm}$, we evaluate both the
multipole moments and the resulting expansions against the real and
imaginary parts of the spherical harmonics $Y_{lm}$ with $m\ge 0$
(noting that $Y_{l0}$ are purely real and that $Y_{l,-m}=(-1)^m
Y^*_{lm}$).  Finally, we truncate the expansion at a predetermined
value $l=l_{max}$, where throughout this paper we use $l_{max}=4$, or
hexadecapole order. Thus, we assume $\rho_{lm}=0$ for all terms with
$l>l_{max}$ when we define the BC's for our field equations.  This is
done for two reasons.  First, the multipole coefficients fall off
steeply at high $l$, so that the higher order multipole make ever
smaller contributions to the field at large separation.  Second,
including higher order multipoles can lead to purely numerical
instabilities in the field solvers for a finite set of Chebyshev
polynomials, since the rapid oscillations with
respect to angle can lead to large gradients in derivative-based
quantities.

We find that that a multipole treatment can lead to significantly
higher accuracy for our boundary solution, at the cost of some
computational efficiency.  To avoid numerical instabilities arising
from quickly growing higher-order multipoles, we employ
underrelaxation during each iteration, updating each field such that
$\psi_{new}=(1-\lambda )\psi_{old}+ \lambda\tilde{\psi}_{new}$, where
$\psi_{old}$ is the field value from the previous iteration, and
$\tilde{\psi}_{new}$ is the new solution found from solving the
elliptic equation.  We find good stability and efficiency by setting
$\lambda=0.5$ initially, and increasing the value to $\lambda=0.5-0.05
\log (\Delta \beta^y)$ with each iteration, where $\Delta \beta^y$ is
the maximum relative change in the y-component of the shift vector
from iteration.  The iteration loop terminates when $\Delta \beta^y<
10^{-9}$, at which point $\lambda \simeq 0.95$, representing very weak
underrelaxation.

Our multipole BC's allow us to calculate the forces on particles that
fall {\it outside} the computational domain directly, since the multipole
expansion for the metric is valid throughout space.    
Indeed, for such particles, we
calculate the BH contribution to the lapse and conformal factor from
Eqs.~(\ref{eq:alphabh}) and (\ref{eq:psibh}), the NS contribution from the
multipole expansions given by Eq.~(\ref{eq:multexp}), and the gradients
of the NS contribution from
\begin{eqnarray}
    \frac{\partial\Phi}{\partial x^i}&=&\partial_i \sum_{l,m}\frac{r^l \rho_{lm}
      Y_{lm}}{r^{2l+1}}\nonumber\\
&=&\sum_{l,m}\rho_{lm}\left[\frac{\partial_i (r^l
        Y_{lm})}{r^{2l+1}}-\frac{(2l+1)x^i r^l Y_{lm}}{r^{2l+3}}\right],
\end{eqnarray}
noting that $\partial r/\partial x^i=x^i/r$.
 
This works directly for the lapse and conformal factor, but the shift vector is
slightly more complicated.  Recall that we have solved elliptic
equations not for $\beta^i$, but for $B_i$ and $\chi$, as defined in
Eq.~(\ref{eq:bchi}), and thus only know the multipole decomposition of
the latter quantities.  In terms of these, the gradient of the shift
is given by
\begin{equation}
\partial_j\beta^i=\frac{7\partial_jB_i-\partial_iB_j-\partial_i\partial_j\chi-
x^k\partial_i\partial_jB_k}{2},
\end{equation}
where we also evaluate terms of the form
\begin{eqnarray}
    \frac{\partial^2\Phi}{\partial x^i\partial x^j}&=&
  \sum_{l,m}\rho_{lm}\left[\frac{\partial_i \partial_j(r^l
        Y_{lm})}{r^{2l+1}}\right.\nonumber\\
&&\left.-\frac{2l+1}{r^{2l+3}}\left(x^i
  \partial_j(r^l Y_{lm})+x^j \partial_i(r^l
  Y_{lm}) \right.\right.\nonumber\\
&&~~~~~~~~~~~~\left.\left.+\delta_{ij}r^lY_{lm}\right)\right.\nonumber\\
&&\left.+\frac{(2l+1)(2l+3)x^ix^jr^lY_{lm}}{r^{2l+5}}\right].
\end{eqnarray}
Since the lapse goes to zero at the horizon, particles approaching it
become frozen in proper time, and cannot penetrate within.

The approach we use has several advantages over the leading order
power-law falloff BC's typically used in BSS and other grid-based
field calculations (e.g., \cite{SU2}).  First, we lose less
information about the source terms by extending to higher order
multipoles (\cite{STU1} include dipole order falloff terms for the
lapse and conformal factor in full GR, while \cite{Oech} include
quadrupole-order terms for these in CF gravity).  Moreover, we avoid a problem
associated with symmetries present in our quasi-equilibrium initial
conditions which are broken during the dynamical evolution.  In
particular, as we show in Appendix~\ref{app:symm}, our
quasiequilibrium configurations can be shown to have a vanishing
monopole contribution to $\beta^x$, and vanishing monopole and dipole
contributions to $\beta^z$.  Once the binary becomes tidally disrupted,
however, we expect the monopole contribution to $\beta^x$ and the
dipole contribution to $\beta^z$ to grow in magnitude ($\beta^z$ may
never have a monopole contribution, since equatorial symmetry holds
for dynamical configurations as well).  While these terms are growing,
we are faced with a situation where the leading-order term may very
well not be the largest magnitude multipole contribution on the
boundary.  Defining a global power-law falloff index to fit the
boundary condition. as in previous treatments, 
is impossible even when the two lowest-order
moments are known, since the index varies with angle.  Instead, our
multipole summation handles this situation naturally, calculating all
low-order moments accurately.

\subsection{SPH Discretization, Computational Domains, and Timestepping}
\label{sec:sph}

Many of the methods used to perform an SPH discretization of the CF
hydrodynamic and field equations are discussed in FGR, so here we
summarize briefly the fundamental aspects of the SPH treatment and
the new features present in our BHNS code.  The neighbor finding
algorithms used in our code are based on routines from StarCrash, a
publicly available, extensively documented Newtonian SPH code, which
can be found online at {\tt
www.astro.northwestern.edu/theory/StarCrash}.

\begin{figure}[ht]
\centering \leavevmode \epsfxsize=\columnwidth \epsfbox{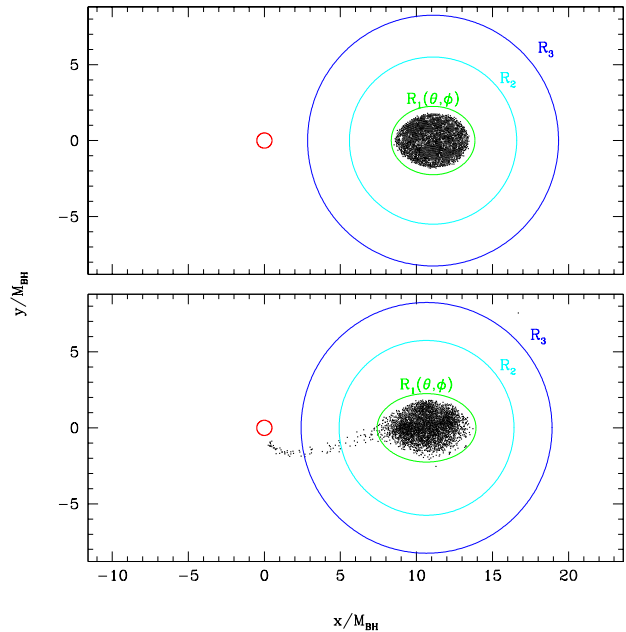}
\caption{A schematic representation of the spectral methods
  computational domains used to solve the NS components of the field
  equations, Eqs.~(\protect\ref{eq:psi})-(\protect\ref{eq:shift}).
  Initially, a triaxial ellipsoid with surface $R_1(\theta,\phi)$ and
  origin at the NS center of mass is fitted around all SPH particles
  for which $(\rho_*)_i>\rho_{crit}$ (top panel).  Two annular
  ``shell-like'' domains with spherical outer boundaries are also laid
  down, with radii $R_2$ and $R_3$, twice and three times the maximum
  value of $R_1$.  During the evolution (bottom panel), we use the
  same procedure to fit $R_1(\theta,\phi)$, keep the value of $R_3$
  fixed, and calculate $R_2$ as the mean of $R_3$ and the maximum of
  $R_1$.  Some particles first leave the innermost domain, and then
  the entire computational volume, particularly those accreted
  by the black hole, shown as a circle of radius $0.5~M_{\rm BH}$
  centered at the origin.}
\label{fig:decomp}
\end{figure}

Motivated by the form of the Lagrangian continuity
equation~(\ref{eq:cont}), we define the mass $m_a$ of each particle,
fixed in time, in terms of the conserved density $\rho_*$, such that
\begin{equation}
(\rho_*)_a=\sum_b m_b W_{ab},\label{eq:rhostarsph}
\end{equation}
where $W_{ab}$ is the ${\cal C}^2$, piecewise, ``$W_4$'' smoothing
kernel function for a pair of particles introduced by \cite{ML}, and used
in FGR and elsewhere.  For each particle, we define a smoothing length
$h_a$, and compute all sums over particles that lie within a sphere of
radius $2h_a$ surrounding each particle 
(we calculate all SPH quantities using a ``gather-scatter''
technique, as described in FGR and the StarCrash documentation).
Smoothing lengths are updated using underrelaxation in order to
maintain a roughly constant number of neighbors for each particle, set
at the beginning of each run.  Each particle is advanced through space
with a velocity $v^i=dx^i/dt$, which we evaluate with a second-order
accurate leapfrog evolution scheme, calculating forces from the Euler
equation~(\ref{eq:dutildedt}) at the half-timestep.  Since the
calculation is adiabatic, the energy equation, Eq.~(\ref{eq:destardt}) is
automatically satisfied when we use the adiabatic EOS,
Eq.~(\ref{eq:polyinit}).  
A typical timestep in our evolution scheme, started with particle
velocities evaluated half a timestep in advance of the particle
positions, involves a number of computational elements.  First, we
advance all particles a full timestep, and re-evaluate the particle
neighbor lists and the SPH expressions for the density of each.  We
then use the SPH form for the density at each particle position to
define the computational domains used by the Lorene field solver,
shown schematically in Fig.~\ref{fig:decomp}.  To do so, we calculate
the position of the NS center of mass from all particles having a
density $(\rho_*)_a > \rho_{crit}$, where $\rho_{crit}$ is a critical
value chosen to encompass the vast majority of particles at the
beginning of a run. Next, we calculate the surface of the innermost
computational domain $R_1(\theta,\phi)$, as the smallest triaxial
ellipsoid, centered on the NS center of mass,
that contains all particles that lie at greater radii from the BH than
the NS center of mass, treating the particles as spheres of radius
$2h_a$.  This is very similar to the technique described in FGR,
except that there we included all particles that passed the density
cut, regardless of which side of the NS they fell within.  Here,
however, the dynamics of the mass transfer are different.  In
equal-mass NSNS binaries, the NS only begin to disrupt at very close
separations, never deviating particularly far from an ellipsoidal
configuration up to the point of merger.  Here, mass transfer is
initially one-sided toward the BH, and the outer half of the NS
remains virtually intact while the inner half becomes deformed by the
tidal gravity of the BH.  We find that our field solver performs best
if we define our elliptical domain based on the profile from the outer
half of the NS, as it can handle without difficulty field sources
located outside the innermost domain, but produces numerical errors if
the density field of the NS drops to zero within the innermost domain.
(see the second panel of
Fig.~\ref{fig:decomp}).  The two outer domains, which have the
topology of spherical shells, are defined initially such that their
outer boundaries are spheres at radii equal to twice and three times
that of the maximum extent of the innermost shell,
i.e. $R_2(\theta,\phi)=2\times \max(R_1); R_3(\theta,\phi)=3\times
\max[R_1]$.  Over time, we hold the outermost boundary fixed at this
radius, and adjust that of the second domain to be the geometric mean
of the outer radius and the maximum value from the inner domain, i.e.,
$R_2=0.5*(R_3+\max [R_1])$ (compare the two panels of
Fig.~\ref{fig:decomp}).

Once the computational domains are defined, we use the techniques of
\cite{BGM} to define a set of ``collocation points'' at which we
compute the local SPH expression for $\rho_*$, $\tilde{u}$, and
$P_*\equiv \kappa\rho_*^{\Gamma}$, noting the latter remains
equivalent to Eq.~(\ref{eq:polytrope}),
for adiabatic evolution and polytropic initial data.  
From these, we calculate all other
hydrodynamic quantities using the Lorene library routines, and solve
the field equations iteratively.  After every iteration of the field
solver, all hydro quantities are updated to reflect the new fields.

Once a convergent solution is found, we must export back all relevant
matter and field terms from the spectral decomposition to the particle
positions.  For particles in the
innermost domain, we evaluate most hydrodynamical terms directly from
the spectral decomposition.  Thus, denoting by ``SB'' those terms
evaluated in the spectral basis and ``SPH'' those quantities defined
only on a particle-by-particle basis, we calculate the Euler equation as
\begin{eqnarray}
\left(\frac{d\tilde{u}_i}{dt}\right)_{in}&=&[\alpha\psi^6]_{\rm
  SB}\left[\frac{\partial_i P}{\rho_*}\right]_{\rm SPH}\nonumber\\ 
&&-\left[\alpha hu^0\partial_i \alpha+\frac{2h\alpha
    (\gamma_n^2-1)}{\gamma_n \psi}\partial_i \psi\right]_{\rm SB}\nonumber\\
&&+\left[\partial_i \beta^j\right]_{\rm SB}[\tilde{u}_j]_{\rm SPH}.
\end{eqnarray}
This approach works in the outermost domains for extrinsic quantities
like $\rho_*$ 
that go to zero smoothly at the surface of the NS matter, but fails
for intrinsic quantities that have discontinuities there, , e.g.,
$u^0$ and $\gamma_n$, since the Chebyshev radial decomposition cannot
describe discontinuous functions.  Instead, we evaluate hydrodynamic
terms for particles in these domains on a particle-by-particle basis,
and evaluate field quantities and derivatives through the spectral
decomposition,
\begin{eqnarray}
\left(\frac{d\tilde{u}_i}{dt}\right)_{out} &=&[\alpha\psi^6]_{\rm
  SB}\left[\frac{\partial_i P}{\rho_*}\right]_{\rm SPH}\nonumber\\ 
&&-[\alpha \partial_i \alpha]_{\rm SB}[hu^0]_{\rm SPH}\nonumber\\
&&+\left[\frac{\alpha}{\psi}\partial_i \psi\right]_{\rm SB}\left[\frac{2h
    (\gamma_n^2-1)}{\gamma_n }\right]_{\rm
  SPH}\nonumber\\ 
&&+\left[\partial_i \beta^j\right]_{\rm SB}[\tilde{u}_j]_{\rm SPH}.
\end{eqnarray}
After calculating the forces for the RHS of the Euler equation, we
advance the velocities from their original half-timestep value forward
to a half-timestep ahead of the positions, and then resolve the field
equations to determine the velocity $v^i$ using the same approach
described above for particles based upon their computational domain,
\begin{equation}
v^i=\left[\frac{1}{\psi^4 u^0}\right]_{\rm SB}[\tilde{u}_i]_{\rm
  SPH}-[\beta^i]_{\rm SB}, 
\end{equation}
in the innermost domain, with $u^0$ evaluated via SPH instead for the
outer ones.
 
\section{Equilibrium models}\label{sec:equil}

The first step in evolving BHNS binaries is the construction of
accurate initial data. In our approach, this requires not only
determining the fields and hydrodynamic quantities within and
surrounding the NS, but also the construction of a relaxed SPH
discretization configuration describing the NS itself.

We take as our starting point data
constructed from the grid-based equilibrium scheme described in BSS.
We modified the scheme of BSS to allow for a conformal background
metric corresponding to a Schwarzschild BH in isotropic
coordinates, which can be adopted more easily for the CF approximation
used here, rather than the Kerr-Schild background used in BSS (see
also TBFS).  To construct an SPH particle decomposition, we first lay
down a hexagonal close-packed lattice of SPH particles over the
Cartesian coordinate volume
where the density of the star is positive.  Tentative particle masses
are assigned to be proportional to the density $\rho_*$, normalized to
match the proper NS mass.  Next, we calculate the SPH value for the
density of each particle, and adjust the masses and smoothing lengths
of each particle until each has approximately the correct number of
neighbors as well as the correct density, to within $\sim 2\%$.  While
the resulting configuration could serve as acceptable initial data, we
can do better by evolving the configuration in the corotating frame
with drag forces applied, to damp away spurious deviations from true
equilibrium.  This also allows us to relax to quasiequilibrium initial
models with binary separations differing by up to $\sim 20\%$ in
either direction using the same initial data from BSS.  Of course, the
new field  solution will be different, reflecting the change in
magnitude of the tidal terms, but we have found that after
approximately 1000 timesteps of relaxed evolution, the overall level
of spurious motion is equivalent.

We used two grid-base datasets to generate our initial data.  For
configuration A, the NS is modeled as a relativistic $\Gamma=1.5$
polytrope, of compaction 
$M_{\rm NS}/R_{\rm NS}=0.042$ (or equivalently, mass
$\bar{M}_{\rm NS}=0.05$ orbiting a BH of mass
$\bar{M}_{\rm BH}=10\bar{M}_{\rm NS}$ at a separation 
$a_0=11.8M_{\rm BH}$.
Configuration B features a NS with the same compaction but
a stiffer EOS, $\Gamma=2$ and
$a_0=11.1M_{\rm BH}$ (where $\bar{M}$ is the dimensionless mass
defined in Sec.~IIIC of BSS).  Note that in these units the maximum
compaction of an isolated NS is $0.214$ and $0.074$ for adiabatic indices
$\Gamma=2$ and $\Gamma=1.5$, respectively.

To convert from the units of BSS to those used here, many quantities
must be linearly rescaled.  In particular, for configuration A,
$\bar{r}=10.5$, and $M_{\rm BH}=4.72$.  Thus all distances should be
multiplied by a factor $\bar{r}/M_{\rm BH}=2.2$ to convert from the
``hatted'' units of BSS to those here expressed in terms of the BH
mass.  Similarly, for configuration B, $\bar{r}=1.32$ and $M_{\rm
  BH}=0.5$, so the rescaling factor is $2.64$.

Both NS models are undercompact compared to the expected physical NS
parameters.  Since our method assumes an extreme mass ratio, we are
limited to low compactness NS models in order to study cases where
tidal disruption occurs outside the ISCO, as can be seen from
Fig.~\ref{fig:qc}.  Thus, while our configurations do not exactly
represent physical parameters expected to be found in BHNS binaries,
they serve as an analogue to binaries containing lower-mass BHs and
more compact NS that will have comparable tidal-disruption radii
located outside the ISCO. In a future work, we will treat more
physically realistic NS compactnesses, as well as NS spins, including
cases for which the tidal-disruption radius is within the ISCO.
 
Below we describe the technique by which we generate our relaxed SPH
initial conditions in Sec.~\ref{sec:relax}, and then show the
comparison between our resulting models and the grid-based data in
Sec.~\ref{sec:quasi}.

\subsection{Relaxation of Initial Data}\label{sec:relax}

When preparing a fluid configuration to be evolved using SPH, it is
generally necessary to use some form of relaxation first.  Otherwise,
numerical deviations from equilibrium present in the discretized
initial configuration will drive the dynamics, leading to a variety of
spurious effects.  Relaxation is easiest to perform for configurations
in which the matter will be stationary in some reference frame, such
as a corotating system, since the spurious component of each
particle's velocity can be easily identified and damped away by a drag
term in the force equation.  This statement holds equally true for
Newtonian and relativistic formalisms, although the latter require a
slightly more complicated numerical treatment, for reasons we discuss
below, primarily due to the presence of velocity-dependent forces as
well as a more complicated set of variables used to define the
equations of motion.

In order to derive the proper equations for a relaxation scheme in a
relativistic setting, it is useful to start with a brief review of how the
process works in Newtonian physics, and then generalize to the
appropriate relativistic equation.  In what follows, we define our
coordinates such that the x-axis corresponds to the line connecting
the centers of mass of the two objects, the y-direction to their
orbital velocity, and the z-direction to the binary's angular velocity.

In Newtonian physics, we have a set of inertial frame evolution equations
\begin{eqnarray}
\frac{d\vec{r}}{dt}&=&\vec{v},\\
\frac{d\vec{v}}{dt}&=&\vec{a},
\end{eqnarray}
where the RHS of each are known functions used to define an initial
condition.  For the case of a corotating equilibrium binary
configuration, these quantities take the form
\begin{eqnarray}
\vec{v}_{eq}&=&\vec{\Omega}\times\vec{r},\\
\vec{a}_{eq}&=&-\Omega^2 \vec{r}_{cyl},
\end{eqnarray}
where we use the subscript ``eq'' to indicate the relaxed value,
and where $\vec{r}_{cyl}$ is the ``cylindrical'' radius.
To evolve the fluid during the relaxation, we shift to the frame in
which the matter is stationary.  Thus we define
\begin{equation}
\vec{V}\equiv \vec{v}-\vec{v}_{eq}=\vec{v}-\vec{\Omega}\times\vec{r},
\end{equation}
so that $\vec{V}=0$ for equilibrium configurations, and evolve
$d\vec{x}/dt=\vec{V}$. We determine $\Omega$ as an eigenvalue from the
condition that the binary center-of-mass separation is already known,
by summing over all of our SPH particles.  Based on symmetry
considerations, only the x-component of the equation yields a
non-trivial result:
\begin{eqnarray}
\sum m_i \left(\frac{d\vec{V}}{dt}\right)^x_i =0&=&\sum_i
m_i \vec{a}^x_i-\sum_i m_i(\vec{a}_{eq})^x_i\nonumber\\&=&\sum_i m_i
\vec{a}^x_i+\sum_i m_i \Omega^2 x_i,
\end{eqnarray}
which implies
\begin{equation}
\Omega=\sqrt{\frac{-\sum_i m_i \vec{a}^x_i}{\sum_i m_i x_i}}.
\end{equation}
We add to the force equation a linear drag term with some
characteristic timescale $\tau$ in order to damp away the spurious
motion in of the initial condition.  The relaxation timescale
should generally be approximately equal to the dynamical time of the
system.  Thus, we evolve
\begin{equation}
\frac{d\vec{V}}{dt}=(\frac{d\vec{v}}{dt}+\Omega^2\vec{r}_{cyl})+\vec{a}_{drag}=
\vec{a}+\Omega^2\vec{r}_{cyl}-\frac{\vec{V}}{\tau}.
\end{equation}
As we approach equilibrium, both the term in parentheses as well as
the drag term separately approach zero.

The relativistic case is slightly more complicated, but we can derive
analogous relativistic expressions for all of our Newtonian ones. 
Our equations of motion now take the form
\begin{eqnarray}
\frac{d\vec{x}}{dt}&=&\vec{v},\\
\frac{d\tilde{u}}{dt}&=&\vec{a},
\end{eqnarray}
where the velocity variables are related by Eq.~(\ref{eq:vofutilde}),
$\vec{v}=\tilde{u}/(\psi^4 u^0)-\vec{\beta}$.  In a corotating frame in
equilibrium, we know that $d(\psi^4u^0)/dt=0$, and will treat the term
as a constant.  The shift vector has the time-dependence of the other
vector quantities.

For a corotating configuration, we have 
\begin{equation}
\vec{v}_{eq}=\vec{\Omega}\times\vec{r},
\end{equation}
and the slightly more complicated Euler equation
\begin{eqnarray}
\vec{a}_{eq}=\frac{d}{dt}[\psi^4u^0(\vec{v}+\vec{\beta})]&=&
\psi^4u^0[\Omega\times(\vec{v}_{eq}+\vec{\beta})]\nonumber\\&=&\!\!
\psi^4u^0[-\Omega^2 \vec{r}_{cyl}+\Omega\times\vec{\beta}]. 
\end{eqnarray}
The matter will be propagated on trajectories with velocity
$\vec{V}=\vec{v}-\vec{v}_{eq}$, just as before.  Now, however, we have
the condition
\begin{equation}
\vec{V}=\vec{v}-\vec{v}_{eq}=
\frac{\tilde{u}}{\psi^4u^0}-\vec{\beta}-\vec{\Omega}\times\vec{r},
\end{equation}
whose time derivative in equilibrium must satisfy the condition
$\vec{a}_{eq}/(\psi^4u^0)-\Omega\times\vec{\beta}+\Omega^2\vec{r}_{cyl}=0$.
In the particle decomposition, we find
\begin{eqnarray}
\sum m_i \left(\frac{dV^x}{dt}\right)_i =0&=&\sum_i
\frac{m_i a^x_i}{(\psi^4u^0)_i}\nonumber\\&+&\sum_i m_i \Omega \beta^y_i +\sum_i
m_i \Omega^2 x_i \nonumber\\
&=&\sum_i \frac{m_i a^x_i}{(\psi^4u^0)_i}\nonumber\\&+&\!\!\!\!\Omega\sum_i m_i \beta^y_i
+\Omega^2 \sum_i m_i x_i ,
\end{eqnarray}
which can be solved for $\Omega$.

To the force equation, we also add a linear drag term with a
characteristic timescale $\tau$, but the drag term must force
$\tilde{u}$ toward its equilibrium value
$\tilde{u}_{eq}=\psi^4u^0(\Omega\times\vec{r}+\vec{\beta})$, rather
than toward zero, so that
\begin{eqnarray}
\frac{d\tilde{u}}{dt}=
\vec{a}-\vec{a}_{eq}-\frac{\tilde{u}-\tilde{u}_{eq}}{\tau}&=&
\vec{a}+\psi^4u^0(\Omega^2\vec{r}_{cyl}-\Omega\times\vec{\beta})\nonumber\\
&-&\frac{\tilde{u}-\psi^4u^0(\Omega\times\vec{r}+\vec{\beta})}{\tau}.
\end{eqnarray}
This equation should, barring any physical instabilities, damp away
spurious motion and produce a corotating equilibrium configuration.

\subsection{Comparison with Other Quasi-Equilibrium Sequences}
\label{sec:quasi}

Our grid-based models, based on the scheme described in BSS but with
isotropic background coordinates, were constructed using $48\times
48\times 24$ grids, with outer boundaries placed $\bar{x}=\pm 3,
\bar{y}=\pm 3, \bar{z}=3$ for the $\Gamma=2$ case, and $\bar{x}=\pm 2,
\bar{y}=\pm 2, \bar{z}=2$ for the $\Gamma=1.5$ case.  SPH
configurations were generated corresponding to these binary
separations, as well as wider and narrower binary separations constructed by
translating the NS to the appropriate position.  The number of
particles used to construct the SPH configurations were $n_p=103,953$
and $n_p=77,908$ for the $\Gamma=2$ and $\Gamma=1.5$ EOS, respectively.
For the $\Gamma=1.5$ EOS configuration only particles of mass
$m_i > 10^{-4} m_{i;max}$ were accepted, where $m_{i;max}$ is the
maximum mass of any SPH particle present.  This mass cut is useful for
eliminating an outermost layer of negligible total mass, which is
typically blown off the surface of the NS anyway by even a tiny amount
of spurious motion resulting from deviations from pure equilibrium in
the initial condition.

Each SPH configuration was relaxed for 1000 time-steps, which
corresponds to $\sim 10 t_D$, a sufficient time given that the initial
models were rather relaxed to begin with.  Parameters for our
grid-based models and the final SPH configurations at the end of
relaxation are listed in Table~\ref{table:ic}.  Models for
$\Gamma=1.5$ are labeled A1 -- A7, while models for $\Gamma=2$ are
labeled B1 -- B4.  As a check, we compare the value of the period
determined during the SPH relaxation process $T$, to the exact
relativistic Kepler relation for a point mass about a BH,
$T_{\rm N}\equiv \sqrt{\tilde{r}^3/M_{\rm BH}}$, and find very
good agreement.  Here the binary separation is measured in {\it areal}
coordinates, whose relation to CF coordinates is given by
(\ref{eq:rbarr}).
 
We note that configurations A1-A3 do not settle down completely during
relaxation.  In each case, the central density of the NS dropped
monotonically as the configuration expanded in the x-direction,
indicating that mass transfer would eventually begin even with drag
forces applied.

\begin{table}[t]
\caption{Parameters for our relaxed initial models.  
   $T_{\rm N}$ is the exact relativistic Keplerian 
  period for a point mass about a BH,
  defined by Eq.~(\protect\ref{eq:rbarr}).}
\begin{tabular}{l|cccc}
Run & $a/M_{\rm BH}$ & $\Omega M_{\rm BH}$ & $T/M_{\rm BH}$ & $T_{\rm
  N}/M_{\rm BH}$\\ 
\colrule\colrule
&\multicolumn{4}{c}{$\Gamma=1.5, q=0.1, {\cal C}=0.042$} \\
\colrule
A1 & 10.438 & 0.0258 & 243.8 & 243.8 \\
A2 & 10.745 & 0.0247 & 254.0 & 253.7 \\
A3 & 11.256 & 0.0232 & 270.6 & 270.3 \\
A4 & 11.513 & 0.0225 & 279.1 & 278.8 \\
A5 & 11.767 & 0.0218 & 287.5 & 287.3 \\
A6 & 12.027 & 0.0212 & 296.5 & 296.1 \\
A7 & 12.791 & 0.0194 & 323.1 & 322.5 \\
\colrule
&\multicolumn{4}{c}{$\Gamma=2, q=0.1, {\cal C}=0.042$} \\
\colrule
B1 & 10.539 & 0.0255 & 246.4 & 247.0 \\
B2 & 10.961 & 0.0242 & 260.2 & 260.7 \\
B3 & 11.093 & 0.0238 & 264.6 & 265.0 \\
B4 & 11.648 & 0.0222 & 283.0 & 283.3 \\
\colrule
\end{tabular}
\label{table:ic}
\end{table}

\begin{figure}[ht]
\centering \leavevmode \epsfxsize=\columnwidth \epsfbox{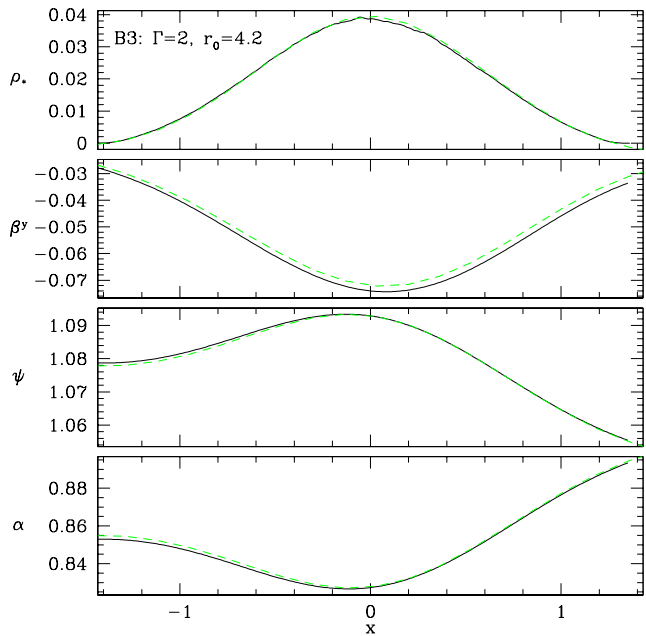}
\caption{From top to bottom, the values of $\rho_*$, $\beta^y$,
  $\psi$, and $\alpha$ along the x-axis for the SPH (solid line) and
  grid-based (dashed line) data representing configuration B3, a NS
  with a $\Gamma=2$ polytropic EOS and an initial binary separation $a_0/M_{\rm
  BH}=11.1$.  The agreement is generally to within $1-2\%$.}
\label{fig:relax1}
\end{figure}

\begin{figure}[ht]
\centering \leavevmode \epsfxsize=\columnwidth \epsfbox{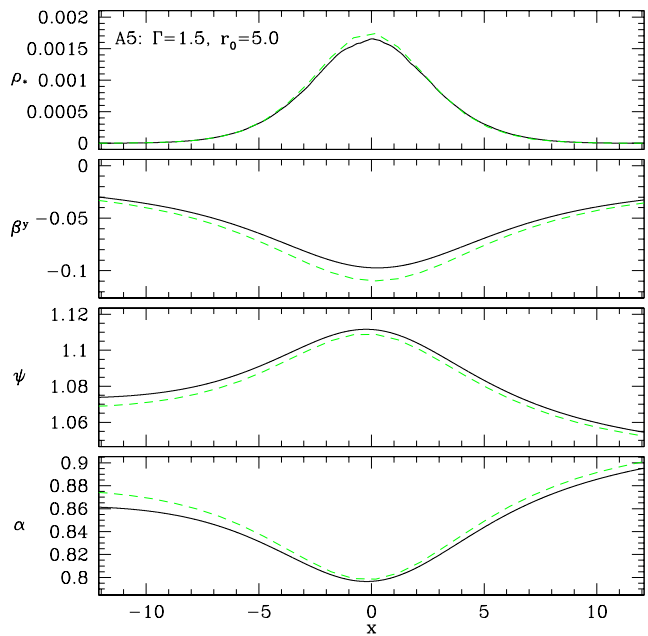}
\caption{The values of $\rho_*$, $\beta^y$,
  $\psi$, and $\alpha$ along the x-axis for the SPH (solid line) and
  grid-based (dashed line) data representing 
configuration A5, a NS with a $\Gamma=1.5$ polytropic EOS and an
  initial binary separation $a_0/M_{\rm BH}=11.8$.  Conventions are the same as
  Fig.~\protect\ref{fig:relax1}.} 
\label{fig:relax2}
\end{figure}

In general, we find very good agreement between the grid-based initial
data and our SPH configurations for stable configurations.  In
Fig.~\ref{fig:relax1} and \ref{fig:relax2} we show a comparison
between the field values and densities from our SPH configuration and
the grid-based data along the x-axis, for configurations A5 and B3.
In both cases the relevant fields agree to generally within about
$1-2\%$.  The only exception is configuration A5, where we find some
disagreement between the two methods on the half of the NS facing the
BH.  The discrepancy is primarily due to the different BC's: the grid
based data imposes a $1/r$ power-law falloff condition on a cube whose
inner edge is located at $\bar{x}=-2$, whereas the multipole solution
used for the SPH configuration is imposed on a spherical boundary at
$\bar{r}=4.5$. Thus, the spectral methods solution integrates over a
much larger volume of space, and allows for higher-order terms in the
field solution at the boundary, which are not insignificant at a
distance corresponding to a few NS radii.  The small disagreement in the
density profile is in part an SPH effect: SPH typically smooths out
the density field over each particle's smoothing length.  Since our
particles are initially equally spaced along a lattice, this length is
$\sim \delta \bar{x}=0.05$, and we cannot fully resolve the sharp
density peak at the NS center.


\begin{figure}[ht]
\centering \leavevmode \epsfxsize=\columnwidth \epsfbox{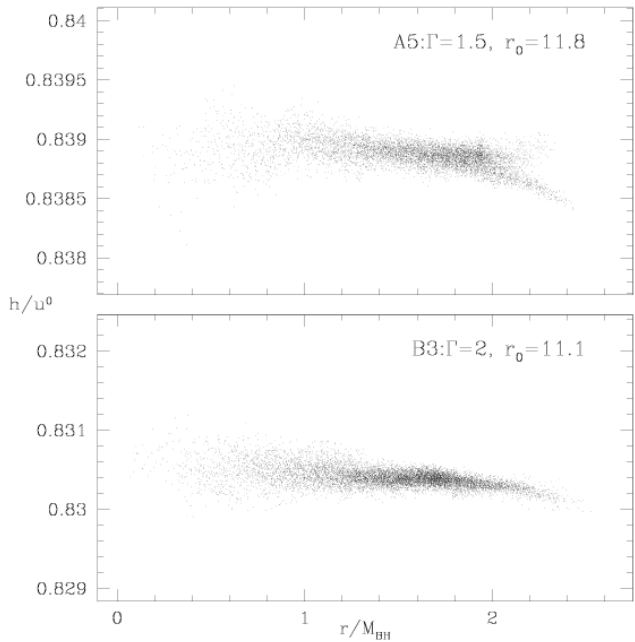}
\caption{Value of the SPH expression for the integrated Euler equation
 constant $h/u^0$ as a function of the particle's radius from the NS
 center of mass for configurations B3
 (top) and A5 (bottom), featuring a NS EOS with $\Gamma=2$ and
 $\Gamma=1.5$, respectively.  Note that the proper value differs
 between the two cases.  The standard deviation in both cases is
 $<0.01\%$, with maximum variation $<0.1\%$.}  
\label{fig:inteuler}
\end{figure}

We can formulate an independent check on the self-consistency of our
initial data by checking how well they satisfy the integrated Euler
equation,
\begin{equation}
\frac{h}{u^0}={\rm constant}.
\end{equation}
This condition is typically used to {\it generate} initial data in
grid-based calculations, e.g., Eq.~(42) of BSS and Eq.~(28) of TBFS, 
but appears nowhere in our
relaxation scheme.  As we show in Appendix~\ref{appendix:invariance},
it can be derived as a {\it consequence} of a relaxed initial
configuration for which the RHS of the Euler
equation~(\ref{eq:dutildedt}) is zero.  In Fig.~\ref{fig:inteuler} we
show on a particle by particle basis the value of $h/u^0$, for
configurations A5 and B3, plotted for clarity against the particle's radial
coordinate position outward from the NS center of mass.  In both
cases, we find the standard deviation from the mean is $< 0.01\%$, and
the maximum discrepancy $<0.1\%$.

\section{Evolution of BHNS binaries}
\label{sec:dynamical}

From our relaxation results, it appears that the tidal
disruption limit for the adopted choices of NS EOS would occur at binary
separations $a_0/M_{\rm BH}\sim 11.0$, in line with the predictions of
Eq.~(\ref{eq:rroche}).  For models with smaller binary separations, we
were unable to find a convergent solution for the configuration with a
stable central density maximum.  These results can be confirmed
through dynamical calculations in the strict CF formalism which ignore all 
energy and angular momentum losses from gravitational
radiation-reaction.  From our discussion in Sec.~\ref{sec:stability},
we might expect the possibility of qualitatively different behavior
for NSs with the stiffer vs. softer EOS evolved from an initial
configuration near the stability limit.  For the softer EOS, we expect
{\it unstable} mass transfer: the NS should disrupt completely once
mass transfer begins.  For the stiffer EOS, we might expect {\it
stable} mass transfer in the strongly viscous regime.  However, as
there is no dissipative mechanism, such as viscosity,
powerful enough to circularize the orbit after the onset of mass
transfer, the picture is considerably more complicated.

All models A (for $\Gamma = 1.5$), or all
models B (for $\Gamma = 2$), describe the same physical binary system, at
different separations representing different moments in its evolution.  
Clearly, for each binary there is only one
correct inspiral history.  In our separate runs we pick up this
history at different points (approximating the orbit as circular), 
which helps to locate the onset
of tidal disruption and analyze the system's dynamical evolution.  
Ideally, we should start an evolution calculation at some large
separation and evolve it forward to complete coalescence, since
the assumption of quasicircularity is violated in an ever increasing
fashion by inspiraling systems. However, even when we augment our CF
equations with a radiation-reaction potential to drive the secular inspiral,
we have neither the time nor the numerical stability to
calculate an evolution for an indefinite time.  Thus, our models
started from different initial separations represent a series of
approximations to the true binary evolution, which illustrate the dynamics
of tidal break-up, and should not be taken as physically distinct
evolutionary paths.  GW energy and angular momentum losses
drive the BHNS binary toward coalescence.  However, the GW timescale
is much slower than the dynamical timescale, so $t_{\rm GW}\ll t_D$, we
expect that radiation-reaction losses will cease to play an
important role in the hydrodynamical evolution once phenomena
associated with the dynamical timescale, such as tidal break-up and
mass loss, begin. 
Nevertheless, our evolution calculations started from outside the stability
limit and including the effects of GW radiation-reaction yield our best models
for the physical evolution of the systems in question. 

\subsection{BHNS mergers with a soft EOS: $\Gamma=1.5$}

As a soft NS EOS, we choose a polytropic model with adiabatic index
$\Gamma=1.5$ (or equivalently, polytropic index $n=2$).  As we see from
Sec.~\ref{sec:relpoly}, a NS of compactness ${\cal C}=0.042$ is
expected to undergo unstable mass transfer in the high viscosity 
limit regardless of the binary mass ratio (and thus in the inviscid
limit as well).  To test out how
well this statement applies in the inviscid limit, which applies to
our calculations (see Tables V and VI of \cite{LomSRS} for numerical
estimates of the viscosity present in a lower-resolution implementation of our
current SPH scheme) as well as to physically reasonable NS, we evolve
binary BHNS configurations from a number of initial separations.  This
also allows us to estimate the critical separation marking the onset
of mass transfer, which according to Eq.~(\ref{eq:rroche}) should be
at $a_R=11 M_{\rm BH}$.

\begin{figure*}[ht]
\centering \leavevmode \epsfxsize=1.8\columnwidth \epsfbox{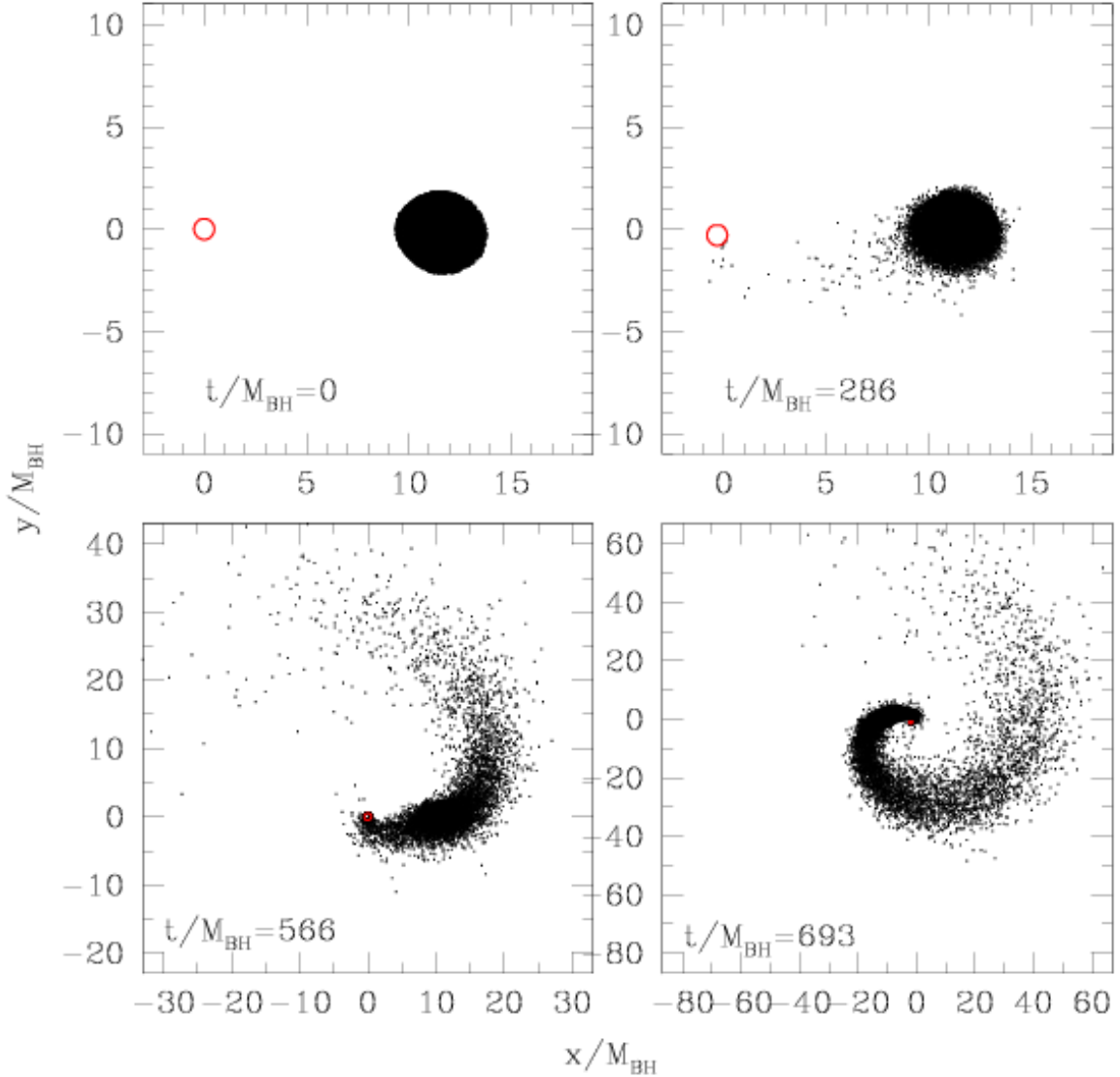}
\caption{SPH particle configurations at times $t=0$, $286$, $566$, and
  $693$, projected into the orbital plane, for run A5 (for which the
  orbital period is $T=287~M_{\rm BH}$).
  These configurations correspond to the initial
  configuration, and after $1$, $2$, and $2.5$ full orbits,
  respectively.  The BH is shown as a circle at a radius $r=0.5M_{\rm BH}$.  
  We see that once mass transfer begins, the NS begins to expand,
  forming a low-density single-armed spiral pattern.  Note that
  particles have different masses, and that those in the center of the
  NS are in most cases significantly more massive than those
  originally in the outer layers.}
\label{fig:xygam15}
\end{figure*}

In Fig.~\ref{fig:xygam15}, we show the evolution of run A5, at an
initial time $t = 0$, corresponding to the initial relaxed
configuration, as well as $t/P=1$, $2$, and $2.5$.  The NS revolves
clockwise around the BH, which is fixed at the origin.  The event
horizon, located at $R_{\rm BH}=0.5 M_{\rm BH}$, is shown as a circle.
In the first plot, the NS has essentially filled its Roche
lobe, and has primary axis ratios
$a_2/a_1\sim a_3/a_1=0.86$.  We note that the figure shows {\it
particle} locations, rather than a surface density representation.  In
fact, particles near the edge of the NS have a density four orders of
magnitude lower than in the NS center.  After a full orbit, we see in
the second panel of Fig.~\ref{fig:xygam15} that the NS has begun to
shed a small amount of mass, which indicates that it is {\it near} the
mass-shedding limit, not necessarily past it.  Our initial SPH
configuration is relaxed to the point where the spurious motion
resulting from deviations from equilibrium is small, but not zero.
Given that the dynamical timescale of the extremely low-density outer
layers of the NS is so long, and the SPH particle masses so small
(roughly proportional to the density), even a tiny error in the
initial data may result in significant spurious velocities in these
layers over time.
For an isolated NS, these particles will remain bound, but the same is
not true for a NS in a binary.  Here, particles that escape the NS
surface will often travel outside the Roche lobe and be lost from the
NS.  As a result, the NS will lose a very small amount of mass and
angular momentum.  By the third panel of Fig.~\ref{fig:xygam15}, the
NS has expanded to the point that matter is now lost through both the
inner and outer Lagrange points.  This leads to the formation of a
stream of matter thrown out into an extended halo around the binary
system, most of which remains bound to the BH.  Meanwhile, there
remains a mass stream of material accreting directly onto the BH.  We
believe that the relativistic nature of the BH gravitational potential
plays an important role in the dynamics of this accretion process.
The matter streaming through the Lagrange point passes sufficiently
close to the BH to fall well within the ISCO on its first passage.  As
a result, most of it accretes {\it directly} onto the BH, rather than
forming a disk.  The instability of orbits near the BH is likely to
play an important role in suppressing the formation of an accretion
disk. We note, however, that our assumption of an extreme mass ratio may
 bias the evolution towards prompt accretion (as does the assumption
 of initial synchronization), since the BH is a fixed target, rather
 than one orbiting the binary center of mass itself.
 Finally, in the last panel, the NS is nearing a state of
complete disruption, and will continue to do so until we can no longer
locate a gravitationally bound object.

\begin{figure}[ht]
\centering \leavevmode \epsfxsize=\columnwidth \epsfbox{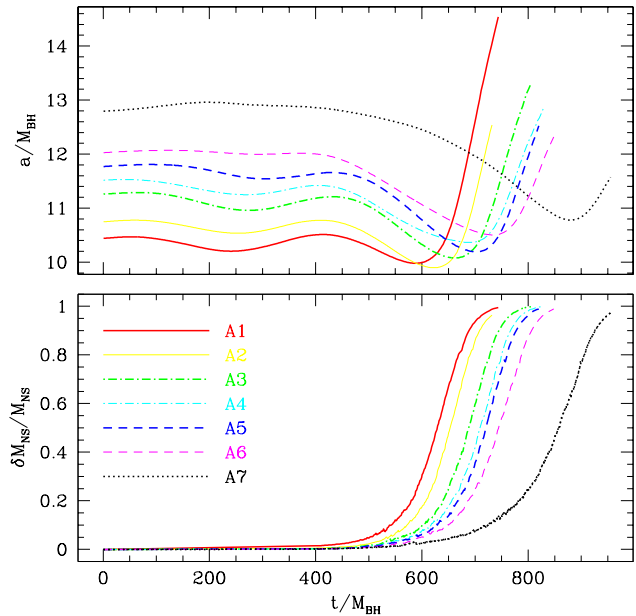}
\caption{Binary separation (top panel) and total mass lost from the NS
(bottom panel) as a function of time for all runs calculated using the
soft NS EOS, runs A1-A7, which have $\Gamma=1.5$.  The properties of
the initial configurations, which differ only in their binary
separation, are shown in Table~\protect\ref{table:ic}.  Here,
particles are defined as ``lost'' when they lie outside the innermost
computational domain around the NS.  In all cases, mass loss never
quenches once it begins, eventually leading to the complete disruption
of the NS.}
\label{fig:rmgam15}
\end{figure}

A similar pattern holds for all runs we performed with the same soft
EOS, regardless of the binary separation.  Due to the nature of the
instability, all models we calculated led to the eventual tidal
disruption of the NS, since there is no stabilizing mechanism to
suppress mass loss once mass transfer begins, no matter how small the
mass-transfer rate.  It does take longer for the NS to be disrupted in
configurations placed at a greater initial binary separation, however.
In the bottom plot of Fig.~\ref{fig:rmgam15}, we show the evolution of
the mass loss over time for all of the configurations using the soft
NS EOS, defined as the total mass that can be found outside the {\it
innermost} computational domain at any given time. We see that in all
cases mass loss is an unstable process, occurring at a rate that grows
extremely rapidly until the
majority of its original mass is no longer bound to the NS. In the
case started from the largest separation, slightly over half of the NS
mass is accreted directly onto the BH or into orbits that lie within
the ISCO, while half is lost outward to form the ``spiral arm''
pattern seen in Fig.~\ref{fig:xygam15}.  This pattern is to be
expected, as the primary response of the NS to losing mass is
expansion.  The inner half of the NS is pushed inward toward the BH,
while the outer half expands outward.  For a synchronized
configuration, we know that the average specific angular momentum of
the NS is generally greater than the specific angular momentum near
the center-of-mass of the NS, since $j\equiv \vec{v}\times\vec{r}\sim
\Omega r^2$ is weighted more heavily by matter on the outside of the
NS.  In response to the expansion of the NS, we expect the orbit to
{\it tighten}, which is confirmed by the numerical results.  In
Fig.~\ref{fig:rmgam15}, we show the binary separation over time for
the runs with a soft NS EOS.  We see that the onset of mass transfer
leads to a rapid decrease in the binary separation, prompting the
explosive mass loss.  Only in the late stages does the remnant of the
NS begin to move back outward, but by then tidal disruption is
inevitable.

\begin{figure}[ht]
\centering \leavevmode \epsfxsize=\columnwidth \epsfbox{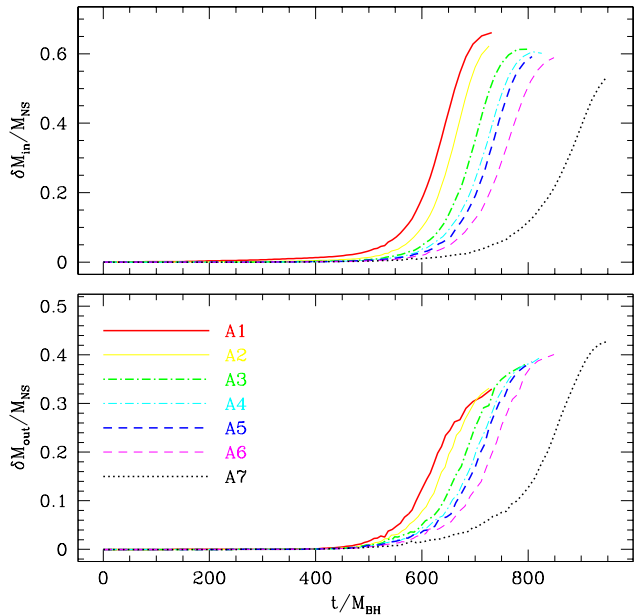}
\caption{Total mass lost inward toward the BH (top panel) and outward
  away from the BH (bottom panel) as a function of time, for the runs
  shown in Fig.~\protect\ref{fig:rmgam15}.  In general, the smaller
  the initial separation, the more mass that falls inward toward the
  BH.  For the runs started from a larger separation, much of the
  initial mass loss results from spurious numerical of low mass
  particles out of the Roche lobe, leading to nearly equal flows
  directed inward and outward.  In contrast, the mass loss for the
  closer cases is in exactly the form expected for Roche-lobe overflow
  through the inner Lagrange point.}
\label{fig:m2gam15}
\end{figure}

As the initial separation is decreased, the qualitative behavior of
the system stays essentially the same, but more of the NS mass ends up
being transferred inward toward the BH. In Fig.~\ref{fig:m2gam15}, we
show the amount of mass lost inward (top panels) and outward (bottom
panels) for all of the runs we performed with the soft EOS.  For
configurations near the stability limit, we find almost twice as much
mass falls inward toward the BH as is expelled outward into the spiral
arm.  As these calculations were performed without radiation-reaction
effects, we expect that including them would tip the balance of the
mass transfer further toward mass accretion onto the BH, since the
binary orbit will be driven by radiation to smaller separations than
we find here.

Unlike the situation found in NSNS binaries, for which $>99\%$ of the
matter typically remains bound to the system, 
we find that a significant
amount of matter is {\it unbound} from the system during all of our
calculations with the soft EOS, representing in each run between
$3-5\%$ of the original mass of the NS.  This fraction is much larger
than that typically found in relativistic calculations of synchronized
binary NS systems \cite{FGR,STU1}, but roughly consistent with
previous results from Newtonian BHNS calculations \cite{LK2,RSW}.
However, we note that this fraction is almost certainly an
overestimate, perhaps greatly so: 
{\it irrotational} configurations suppress the
amount of mass that will become unbound (see, e.g., \cite{FR3} or FGR for a
similar argument with respect to NSNS mergers) .  The total angular momentum
of an irrotational configuration is less than that of a synchronized
configuration, and the decrease in specific angular momentum is
largest at the outer edge of the NS.  Thus, we expect that while some
mass may still be unbound when the NS is initially irrotational, the
amount may be significantly less.

\subsection{BHNS mergers with a stiffer EOS: $\Gamma=2$}\label{sec:gam2}

To study a configuration that would be predicted to undergo {\it
stable} mass transfer in the classical conservative quasiequilibrium 
scenario, described in Sec.~\ref{sec:stability}, we
model the NS with a stiffer, $\Gamma=2$ polytropic EOS, which has
been used in numerous studies
as a first approximation to a stiff NS EOS.  We note that
in Newtonian physics, this is the critical polytropic index for which
an isolated NS has a radius independent of its mass.  In relativistic gravity,
self-gravity effects cause the NS radius to increase as the NS mass
decreases.

\subsubsection{Evolution without GW radiation-reaction}

\begin{figure}[ht]
\centering \leavevmode \epsfxsize=\columnwidth \epsfbox{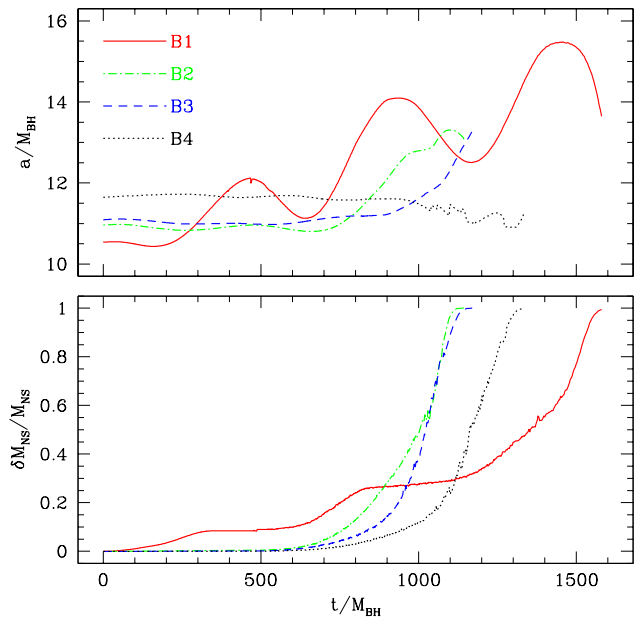}
\caption{Binary separation (top panel) and total mass lost from the NS
(bottom panel) as a function of time for the runs calculated using the
stiff NS EOS without GW radiation-reaction, runs B1-B4, which have
$\Gamma=2$.  Conventions are as in Fig.~\ref{fig:rmgam15}.  For run
B1, we see periodic mass loss, as the NS gets kicked into an
elliptical orbit, losing mass during every periastron passage.  Runs
B2 and B3 are similar to those with the softer EOS, as the NS gets
disrupted during the first mass loss phase.  Run B4 is essentially
stable, only showing the long-term effects of numerical diffusion of
particles.}
\label{fig:rmgam2}
\end{figure}

In Fig.~\ref{fig:rmgam2} we show the binary separation (top panels)
and NS mass fraction lost (bottom panels) as a function of time for
runs B1-B4.  We find three qualitatively different behaviors over
time, which we will describe in turn.  We note first that the tidal
limit separation is not drastically different for this EOS compared to
the softer EOS.  Both cases show very good agreement with the
classical Roche limit result, as we would expect.

\begin{figure}[ht]
\centering \leavevmode \epsfxsize=\columnwidth \epsfbox{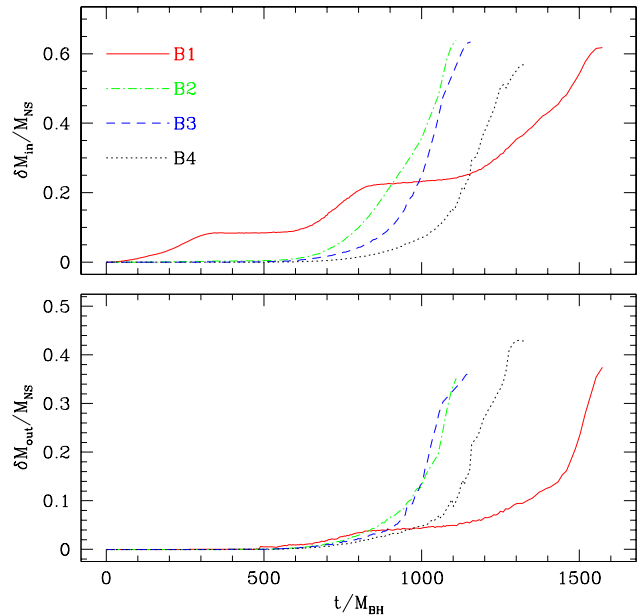}
\caption{Total mass lost inward toward the BH (top panel) and outward
  away from the BH (bottom panel) as a function of time, for the runs
  shown in Fig.~\protect\ref{fig:rmgam2}. Conventions are as in
  Fig.~\ref{fig:m2gam15}. Approximately $50\%$ more
  mass is lost inward for runs B2 and B3, started from near the
  stability limit, whereas for run B1, marked by an elliptical orbit
  and periodic mass transfer, nearly $4$ times much mass is lost
  toward the BH.}
\label{fig:m2gam2}
\end{figure}

Run B4 is completely stable for approximately three orbits.
Eventually, the NS begins to lose a small amount of mass, which gives
us an estimate for the length of time over which the code can reliably
maintain an equilibrium configuration in the absence of relaxation
($t/T\sim 3$).
We note from Fig.~\ref{fig:m2gam2} that the mass loss from the NS is
almost evenly divided between a component moving toward the BH and the
component directed away, indicating that the mass loss is a numerical
artifact caused by particles near the edge of the NS diffusing outside
the Roche lobe over time.

Runs B2 and B3 are started from a binary separation approximately
equal to the mass-shedding limit.  In both cases, the NS makes
approximately one orbit after mass transfer begins without any
appreciable change in the binary separation.  By this point,
approximately $10\%$ of the NS mass is stripped away, and the binary
begins to move outward.  As we saw in the case of the softer EOS, the
mass loss is unstable and the mass-transfer rate grows 
rapidly.  In both cases, the NS is tidally disrupted before the
mass transfer halts.  By the time the NS is disrupted, approximately
$60\%$ of the original mass has fallen inward toward the BH, as we see
from Fig.~\ref{fig:m2gam2}.  Of the matter shed outward, approximately
$0.075 M_{\rm NS}$ is unbound from the system.  This result fits in
with the general picture derived from NSNS binaries, in which mass
shedding outward is more efficient for stiffer EOS (see, e.g.,
\cite{FR2} and references therein).

The evolution of the NSs in runs B2 and B3 started from initial
conditions differing only in their binary separations (run B3 starts
from $0.5\%$ further out), which are nearly equal to the
mass-shedding limit.  We find that the closer the NS is
when mass transfer begins, the more mass is lost inward toward the BH
relative to that lost outward, until the NS expands to
the point that it greatly overfills its Roche lobe and mass loss
becomes more isotropic around the axis describing the NS velocity.

These observations go a great deal toward clarifying the evolution
seen in run B1, the only case in which the mass transfer was periodic,
rather than continuous.  Here the initial configuration places the NS
{\it within} the mass-shedding limit, as GW losses would be expected
to do for physical BHNS systems.  Indeed,runs B3a and B3b, described
in Sec.~\ref{sec:radreac}, which include the effects of radiation
reaction, are driven to approximately this binary separation by GW
radiation-reaction energy and angular momentum losses.
When mass transfer begins, it occurs
only toward the BH, which causes the NS to move outward.  This
expansion in the orbit, which happens with no outwardly directed
stream to counter the outward acceleration, occurs sufficiently fast that
the Roche-lobe expansion is enough to quench mass transfer.
At this point, the NS is on a continuously expanding eccentric orbit
($e\sim 0.1$), whose apocenter lies outside the Roche limit and
pericenter within it.  As the NS crosses over the new mass-shedding
limit, mass transfer begins again, pushing the NS onto an even wider
eccentric orbit, similar to the pattern seen in \cite{RSW} for NS with
a stiffer EOS.

This scenario is ostensibly similar to that proposed by \cite{DLK},
but with one crucial difference.  In both cases, mass transfer forces the
NS onto a highly elliptical orbit.  Here, however, mass transfer
occurs during {\it every} periastron passage, whereas in their model
the NS can be kicked into such a widely separated orbit that GW
radiation-reaction must drive the NS back toward the mass-shedding point.
We believe that the assumptions that go into the latter model lead to
this unphysical result.  In \cite{DLK}, it is assumed that the NS
loses a specified amount of mass during mass-transfer events but
recovers half of its angular momentum during the next half orbital
period.  This can lead to a discontinuous evolution of the binary
separation.  Here, we see that once mass transfer ceases, the NS will
follow a nearly unperturbed elliptical orbit, with essentially no
change in its orbital angular momentum.  
Mass transfer must resume when the orbit
crosses this same point as it approaches pericenter during the next
passage.

\subsubsection{Mergers including GW radiation-reaction}\label{sec:radreac}

While evolution calculations lacking GW radiation-reaction terms can be
useful for studying the processes that control the moment to moment
dynamical evolution of the system, we know that the effects of
radiation-reaction must play an important role in the secular
dynamics of the system.  Indeed, given the potentially unstable nature
of mass loss, we expect the inwardly directed component of the NS
velocity to be critical.  Since the mass transfer leads not to an
instantaneous change in the NS velocity, but rather in its
acceleration, there will be a time period immediately after the onset
of mass transfer during which the NS loses mass while falling further
{\it inward}. All the while, the inner Lagrange point will move
further within the NS, regardless as to how its radius adjusts on the
dynamical timescale.

To model radiation-reaction, we add a damping force to the material,
representing the lowest-order quadrupole contribution to the radiation
reaction potential.  Thus, we add to the RHS of Eq.~(\ref{eq:dutildedt}) an
acceleration term of the form
\begin{equation}\label{eq:rrforce}
a_{i:reac}=N^2hu^0\partial_i \chi,
\end{equation}
where $\chi$ is a quadrupole radiation-reaction potential (see
Sec.~36.1 of \cite{MTW}), defined here as 
\begin{equation}
\chi=\frac{1}{5}x^k x^l Q^{[5]}_{kl},
\end{equation}
in terms of the fifth time derivative of the quadrupole moment, 
\begin{equation}
Q_{kl}=STF\left[\int \rho_{\rm ADM} x_k x_l d^3 x\right].
\end{equation}
Here $\rho_{\rm ADM}$ is the quantity that can be integrated to give the
matter contribution to the ADM mass $\rho_{\rm ADM}\equiv \psi^5 E$,
which appears in the field equation for $\psi$, Eq.~(\ref{eq:psi}).
Following the argument found in Sec. IIIa of FGR, we evaluate
the fifth time derivative of the quadrupole moment in terms of the
expression for the first time derivative in the limit that it is
dominated by terms representing orbital motion and not changes in the
density with time (i.e., $d\rho_{\rm ADM}/dt$ is negligible),  
\begin{equation}
(\dot{Q}_1)_{kl}=STF\left[\int \rho_{\rm ADM} (x_k v_l+x_l v_k) d^3 x \right],
\end{equation}
where ``STF'' means the symmetric trace-free component of the tensor.
We then assume that further time derivatives result purely from the
orbital motion, and approximate 
\begin{equation}
Q^{[5]}_{kl}\approx 16\omega^4 \dot{Q}_{kl},
\end{equation}
where the angular velocity $\omega$ is taken as the ratio
of the angular momentum to the moment of inertia, 
\begin{equation}
\omega=\frac{\sum_a m_a (x_a (v_y)_a - y_a (v_x)_a)}{\sum_a m_a (x_a^2+y_a^2)}.
\end{equation}
While the resulting radiation-reaction force will differ slightly from
the true quadrupole expression found by taking the exact time
derivatives, it is sufficiently accurate for our purposes here,
generally within $10\%$.  It is
important to remember that the radiation-reaction force drives the
binary inward on a secular timescale $t_{\rm GW}$, but plays almost no role
once effects occurring on the much more rapid dynamical timescale
become dominant.

\begin{figure*}[ht]
\centering \leavevmode \epsfxsize=1.8\columnwidth \epsfbox{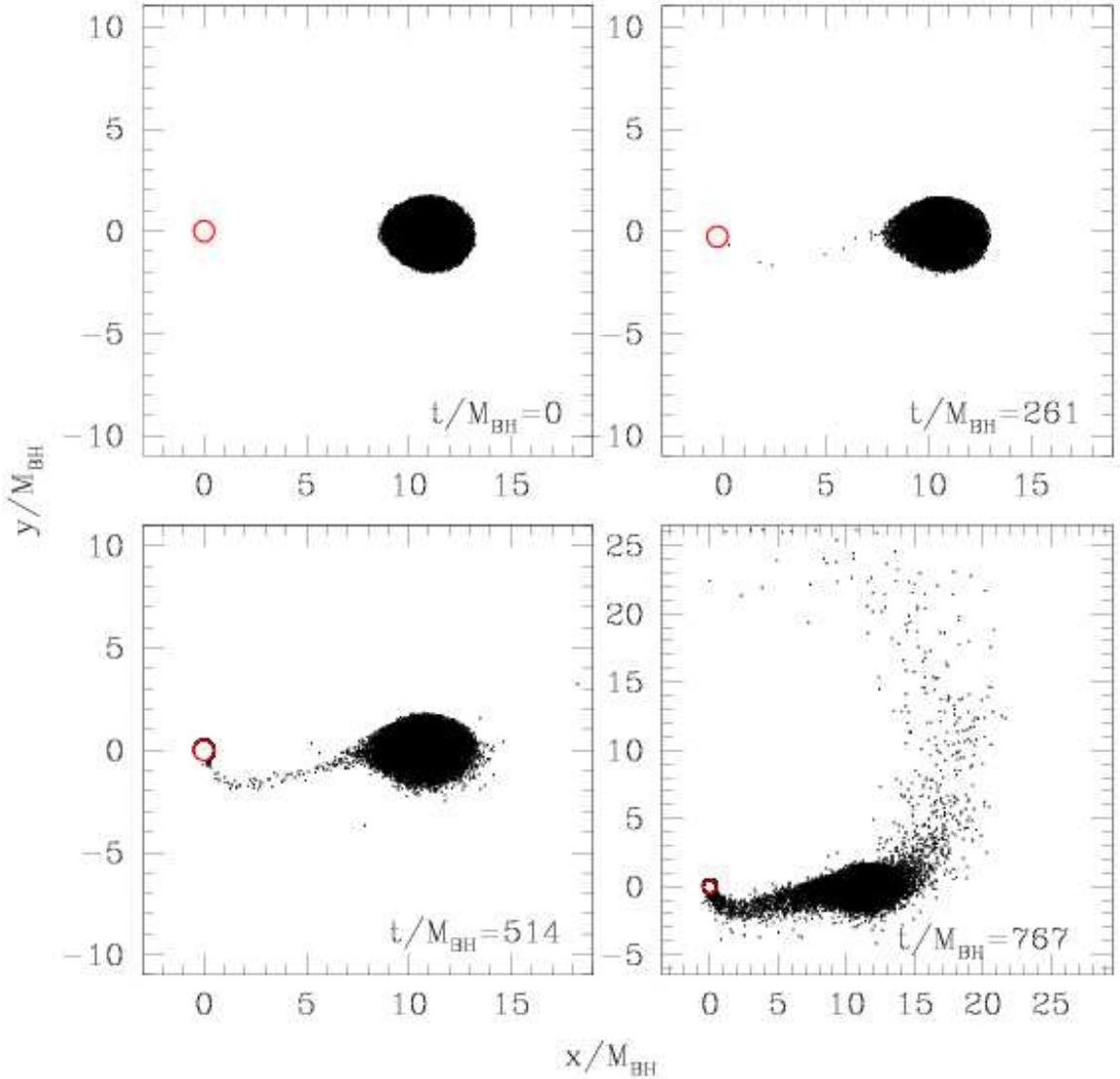}
\caption{SPH particle configurations at times $t/M_{\rm BH}=0$, $261$, $514$,
  and $767$, projected into the orbital plane, for run B3a, which
  includes the dissipative effects of radiation-reaction (for this
  configuration, $T=265 M_{\rm BH}$).  In terms of the
  binary orbit, these correspond to the initial configuration and 
  $1$, $2$, and $3$ full orbits, respectively. Conventions are as in
  Fig.~\ref{fig:xygam15}. 
  Here, we see that radiation-reaction initially drives only an
  inwardly directed mass-transfer stream onto the BH, until somewhere
  after $t/M_{\rm BH}=700$, when the expansion of the NS becomes unstable and
  tidal disruption occurs.}
\label{fig:xygam2}
\end{figure*}

To test how radiation-reaction effects change the scenario
described above, we calculated two
runs that included radiation-reaction terms.  Both took as initial
data the configuration used also for run B3, using the stiffer
$\Gamma=2$ NS EOS, placed nearly at the stability limit.  For run B3a,
we added a radiation-reaction acceleration term 
described by Eq.~(\ref{eq:rrforce}); for run B3b, we doubled the
magnitude of this force.  The evolution of the former, run B3a, is
shown in Fig~\ref{fig:xygam2}.  Mass transfer occurs in a
well-defined stream for the first two orbital periods, until the
expansion of the NS eventually drives the rapid tidal disruption of
the NS.

\begin{figure}[ht]
\centering \leavevmode \epsfxsize=\columnwidth \epsfbox{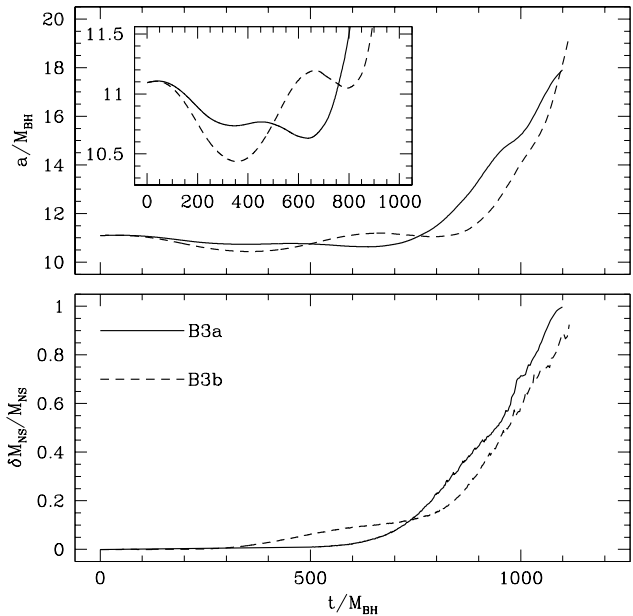}
\caption{Binary separation (top panel) and total mass lost from the NS
(bottom panel) as a function of time for the runs calculated using the
stiff NS EOS and dissipative GW radiation-reaction effects, runs B3a
and B3b, which have $\Gamma=2$.  Conventions are as in
Fig.~\ref{fig:rmgam15}.  The only difference between the two runs is
the magnitude of the radiation-reaction effects; we use the quadrupole
order form in run B3a, and double its strength for run B3b.  We find
that doubling the radiative drag force forces the binary to a smaller
separation, leading to a larger initial burst of mass loss and a more
rapid increase in the binary separation.  As a result, the system
tidally disrupts slower than the case shown in run B3a.}
\label{fig:rmrr}
\end{figure}

\begin{figure}[ht]
\centering \leavevmode \epsfxsize=\columnwidth \epsfbox{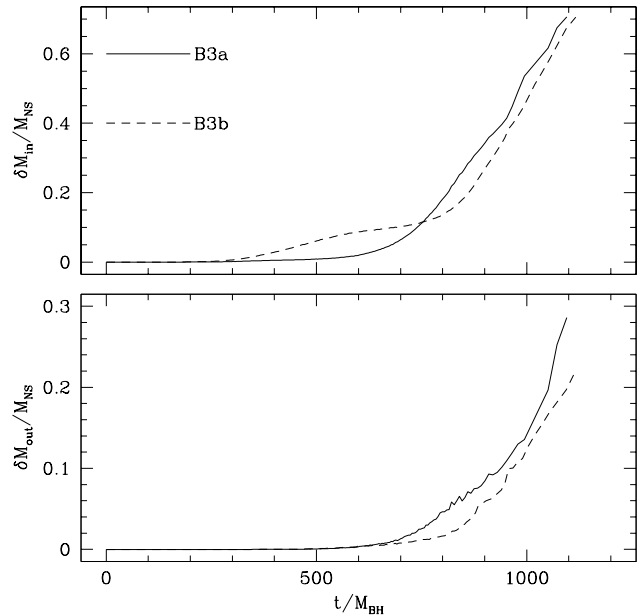}
\caption{Total mass lost inward toward the BH (top panel) and outward
  away from the BH (bottom panel) as a function of time, for the runs
  shown in Fig.~\protect\ref{fig:rmgam2}. Conventions are as in
  Fig.~\ref{fig:m2gam15}.  Initially, in run B3b with twice the
  physical GW radiation-reaction force applied, all mass lost in run
  B3b is directed inward.  This leads to a rapid increase in the
  binary separation and slows the growth of the mass-transfer rate.
  In contrast, in run B3a, mass loss is more evenly balanced between
  inwardly and outwardly directed flows, and the NS disrupts more quickly.}
\label{fig:m2rr}
\end{figure}

In Fig.~\ref{fig:rmrr}, we show the evolution of the
binary separation (top panels) and NS mass loss (bottom panels) for
runs B3a and B3b.  We find that the stronger the
radiation-reaction losses, the greater the initial mass loss from the
NS, since the first passage within the mass-shedding limit takes it
closer to the BH.  This in turn drives the NS back outward, quenching
the mass loss
temporarily until the next periastron passage, after which the NS
tidally disrupts.  From Fig.~\ref{fig:m2rr}, we see that stronger
radiation-reaction losses favors a larger amount of mass lost inward
onto the BH (and thus less outward into a disk), 
as one would expect. Thus, we conclude that the inclusion
of GW radiation-reaction terms have the effect of increasing the
chance that some fraction of the original NS mass will remain bound
after an initial phase of mass loss, since the NS rebounds more
sharply outward than for cases in which radiation-reaction losses are
ignored. Tidal disruption, while still seemingly inevitable, also
occurs at a greater distance from the BH.

\begin{figure}[ht]
\centering \leavevmode \epsfxsize=\columnwidth \epsfbox{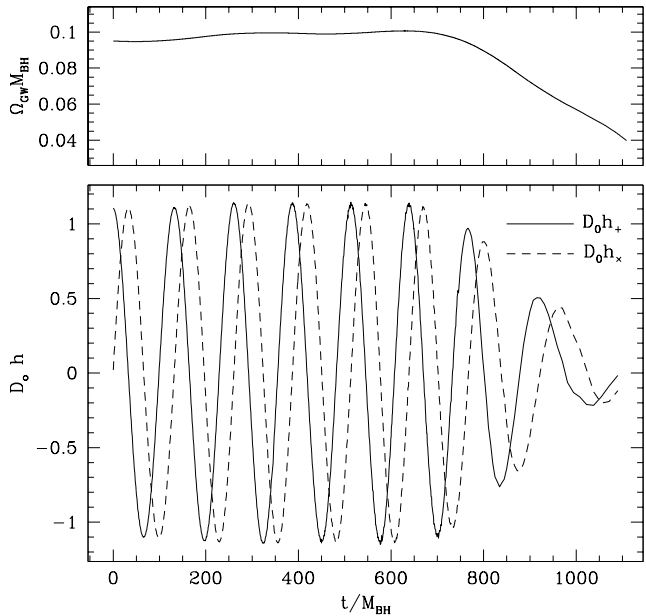}
\caption{Gravity wave emission angular frequency (top panel) and
  waveform in both polarizations, $h_+$ (solid curve) and $h_\times$
  (dashed curve), defined by
  Eqs.~(\protect\ref{eq:hplus}) and (\protect\ref{eq:htimes}), for the
  merger calculated in run B3b.  Initially, we see
  conclusion of the standard ``chirp'' signal, in which the binary
  separation decreases while the GW amplitude and frequency
  increases, all of which happens extremely gradually on the secular GW
  timescale.  This lasts until the onset of mass transfer, at which
  point we encounter a much more rapid ``reverse chirp'', as the GW
  amplitude and frequency rapidly decrease while the NS is tidally disrupted.}
\label{fig:gwrr}
\end{figure}

In the bottom panel of Fig.~\ref{fig:gwrr}, we show the gravity wave
signal produced in run B3a, in both polarizations.  The two components
are defined by the familiar relations
\begin{eqnarray}
D_oh_{+}&=&\ddot{Q}_{xx}-\ddot{Q}_{yy},\label{eq:hplus}\\
D_oh_\times&=&2\ddot{Q}_{xy},\label{eq:htimes}
\end{eqnarray}
where $D_o$ is the distance from the observer to the binary.  The
corresponding angular frequency of the GW signal, approximately equal
to twice the orbital angular frequency, is shown in the top panel of
the figure.  Prior to disruption, the GW waveform takes the
classic point-mass form, with a steadily but extremely slowly
increasing amplitude and
frequency (relativistic and finite-size affects cause minor
deviations from the point-mass form; see \cite{FGRT} for a
discussion).  Once the mass transfer begins, however, the frequency
reaches a maximum and begins to decrease quickly, as does the
amplitude.  In general, if the NS does transfer sufficient mass to
survive the initial infall, we expect this decrease in frequency and
amplitude until the signal can no longer be reliably observed.  Should
the orbit be elliptical, as we found for run B1, this will show up in
the GW waveform as well.

As matter accretes onto the BH, it will excite quasi-normal ringing modes that
could in theory be visible in the gravitational wave signal.  
Unfortunately, our numerical approach limits our ability to determine
the gravitational wave signal we expect from these modes.  Indeed,
such a calculation would require a dynamical treatment of the
spacetime very near the BH, whereas in our calculation, the key
physics for ringdown occurs in the asymptotic region located outside our
computational domain, where the BH remains
stationary.  To evaluate quasi-normal mode ringing, it will be
necessary to relax the CF approximation, in which no gravitational
radiation is generated, and to evolve the fields everywhere.
Additionally, the use of a lapse function that penetrates the horizon
will be crucial for studying the problem self-consistently.  We plan
to study these issues in future work.

\section{Summary and Discussion}\label{sec:discussion}

We have performed relativistic calculations of BHNS mergers in CF
gravitation, in the limit that the BH is much more massive than the
NS.  These calculations mark the first
time that a relativistic treatment has been applied both to the
self-gravity of the NS as well as the BH spacetime.

For systems studied here in which the onset of tidal disruption occurs
outside the ISCO, previously proposed analytical models do not properly
describe all the complex tidal phenomena
that pertain to the evolution.  In all the runs we calculated, mass
transfer plays a leading role in determining the dynamics of the
system.  In general, mass transfer in a stream directed toward the BH
causes the orbital separation to increase, in many cases quite
dramatically.  Mass transfer also causes the NS radius to expand on
the dynamical timescale, with {\it relativistic} self-gravity effects
leading to a more rapid expansion than is seen in Newtonian gravity
for a given NS EOS.  As a result, mass transfer is significantly more
unstable in relativistic gravity.  We conclude immediately that
previous models of mass transfer in compact object binaries that
assume the orbit remains quasi-circular \cite{CE,PZ} are not
applicable here.  Furthermore, the model put forward in \cite{DLK}
also seems to be insufficient, in that the orbital parameters evolve
discontinuously from one orbit to the next (as a result of angular
momentum being added to the NS while its mass is held fixed).  
We find instead that if some
remnant of the NS survives the initial burst of mass loss, it can end
up on an elliptical orbit that takes it back outside its mass-shedding 
separation.  During {\it every} successive periastron passage,
however, more mass will be lost, eventually leading to the complete
disruption of the NS. 

As the NS expands during mass loss, it eventually loses mass outward
as well as inward, so long as the plunge does not take it too far
within the ISCO, leading to a prompt plunge onto the BH.  While the
majority of matter released through the outer Lagrange point remains
bound to the BH in the former case, a significant fraction is
ejected with sufficient velocity to become unbound from the system
completely, approximately $5-7\%$ for the $\Gamma=2$ EOS we
considered.  This mass loss also limits the radial expansion of the
orbit, and as a result we find in many of our calculations that mass
transfer is never quenched once it begins.  Even though the NS moves
outward, it persists in configurations for which the Roche lobe lies
within the star, and mass loss continues until the NS is completely
disrupted.

We plan to improve our simulations and relax several approximations in
the near future.  In particular we plan to adopt the astrophysically
more realistic {\it irrotational} initial configuration of TBFS
instead of the corotating configurations used here.  Relativistic NSNS
\cite{FGR,SU1} and Newtonian BHNS \cite{RSW} calculations have shown
that for these cases the amount of matter ejected from the binary
system is expected to be significantly smaller, since the material on
the outer edge of the NS has significantly less specific angular
momentum.  Thus, while we expect that BHNS mergers will expel more
material than in the case of NSNSs, in which the binary components are
more comparable in mass, we would assume that the ejected fractions
found here are overestimates.

Calculating mergers using irrotational NSs should also increase the
probability that escaping matter forms an accretion disk around
the BH, even if that disk is short-lived (see, e.g.,
\cite{Ross05}).  We typically 
found in our calculations here that most of the matter transferred
toward the BH ends up accreting onto it directly, since the specific
angular momentum is not sufficient to create a disk.  Even though an
irrotational NS has less total angular momentum than a corotating one,
the matter on the inner edge has a higher specific angular momentum,
and is more likely to orbit around the BH rather than infall directly.
In addition, prompt accretion of matter may also be inhibited slightly by a
moving BH orbiting the binary center of mass, whereas in our assumption
of an extreme mass ratio, the BH position is fixed.

In our future work, we will test out how shock heating in the
accretion disk affects its evolution, and determine if there are
cases in which feedback onto the NS will affect its future evolution
as well.  By including a
relativistic artificial viscosity treatment, we will follow the
thermodynamic evolution of the disk, as well as that of the NS and any
outward mass loss.
Of course, to investigate BHNS mergers fully and accurately, 
we will ultimately have to abandon
the assumptions underlying the CF metric as well.  While our
description of an isolated Schwarzschild BH is exact, the BH lapse
goes to zero just outside the event horizon.  
This causes matter to ``pile up'' around the BH,
since the proper time ceases to advance in this region.  While
this poses no difficulty for determining the fate of the material,
which will clearly be accreted, it does act as a computationally
challenging ridge of growing mass concentration
as material is transferred continuously toward the BH.

These calculations adopted ``undercompact'' NSs because we
are interested in systems that disrupt outside the ISCO while
restricting our attention to the case of extreme mass ratios.  By
performing dynamical evolutions using more compact configurations, we
will investigate the transition to the opposite case, in which plunge
begins while the NS is still completely bound.  Based upon our
results, we expect that that in some cases, the mass-transfer rate may
prove sufficient to kick the core of the NS back out to a wider,
highly elliptical orbit.  The phase space for which this will occur, in
terms of the NS EOS and binary mass ratio, remains poorly understood,
and may not conform to simple analytic estimates, which have
difficulty describing the unstable processes that characterize the
merger.  In \cite{RSW}, it was found for a quasi-Newtonian potential
that for a mass ratio $q=0.1$, a NS described by a $\Gamma=2$ EOS was
disrupted during the initial passage, whereas one with a $\Gamma=3$
EOS led to a punctuated mass-transfer scenario similar to the one we
describe for run B1 above in Sec.~\ref{sec:radreac}.  Probing the
assumed forms of the NS
EOS that lead to complete disruption versus survival of a remnant NS
core may prove to be a crucial diagnostic tool for determining the
true physical NS EOS.

In many prior works, it has been assumed that so long as
$a_{ISCO}<a_{Roche}$, the NS will necessarily be tidally disrupted,
with some of its mass deposited into a disk around the BH.  According to
\cite{Miller}, this may not be true.  
Instead, the plunge may start well outside of the ISCO,
since, based on the properties of quasiequilibrium models, the
inspiral time scale may approach the orbital time scale already well
outside of the ISCO (so that the transition through the ISCO is
nearly dynamical rather than adiabatic).  It is unclear exactly what
role the spin of the BH will play in this process, though it appears
that a prompt merger is more likely for a Schwarzschild BH than a
spinning Kerr BH \cite{PSBH}, where the latter case is favored by
binary evolution calculations \cite{PRH,OKKB}.

Two different factors can mitigate
a prompt merger, however, 
and will need to be investigated numerically in more
detail.  First, the onset of a ``plunge'' also marks the point at
which we assume the quasi-equilibrium formulation to break down.  From
that point onward, the NS will no longer follow the trajectory
predicted by quasiequilibrium results, and may therefore not plunge as
fast as quasiequilibrium models would predict (in fact, these
predictions provide completely unphysical overestimates of the infall
velocity at the ISCO itself).  Perhaps more importantly, there is no
guarantee that a plunge phase will lead to the entire NS being
swallowed by the BH \cite{Ross05}.  Angular momentum can be
transferred outward within the NS on something approximating the
dynamical timescale, and it is possible that some fraction of the
mass, perhaps a significant fraction, will survive the initial plunge
phase.  Using the techniques developed here, and initial
configurations taken from TBFS, we will study how varying the NS spin
and compaction affect the final fate of the NS.

It will be necessary to perform
relativistic merger calculations in full generality, in order to drop the
assumption that the binary mass ratio is extreme.  We are currently
constructing such quasiequilibrium data, which will then serve as
initial data for dynamical simulations \cite{TBFS2}.  
We expect that these dynamical
simulations will be plagued by the same difficulties encountered in
dynamical simulations of BHBH binaries, 
and therefore anticipate that this will be a very challenging project.

Until that point, however, a great deal can be
accomplished.  First and foremost would be to identify the boundaries
in phase space that separate qualitatively different phenomena that
can occur during a BHNS merger.  The most obvious categories would be
prompt merger, prompt tidal disruption leading to an accretion
disk, or a period of periodic mass-transfer bursts, if the latter
does occur at all.
The former distinction should prove useful for
understanding any potential X-ray/gamma-ray emission from these
systems, and aid in our understanding of short gamma-ray bursts like
the recently observed GRB 050509b \cite{SGRB,Gehrels}, GRB 050709
\cite{Covino,Fox}, GRB 050724\cite{Berger}, and GRB 050813\cite{050813}.

\begin{acknowledgments}
JAF is supported by an NSF Astronomy and Astrophysics Postdoctoral
Fellowship under award AST-0401533.  TWB gratefully acknowledges
support from the J.~S.~Guggenheim Memorial Foundation.  This work was
supported in part by NSF grants PHY-0205155 and PHY-0345151 and NASA
Grant NNG04GK54G to the University of Illinois, NSF Grant PHY-0139907
to Bowdoin College, and PHY-0245028 to Northwestern University.
\end{acknowledgments}

\appendix

\section{Mass transfer in the viscous regime}\label{appendix:masstransfer}

Below, we derive a formalism that can be used to describe conservative
quasiequilibrium mass transfer, for those cases where the viscosity is
high, and the binary orbit remains circular during the mass-transfer process.
For typical (high-mass) NSs, this regime does not apply 
(\cite{BC,Koch}; see also
Fig.~\ref{fig:qc} above), but can apply to low-mass NSs, WDs, and main
sequence stars orbiting BHs.  We nevertheless refer to the star as a
NS below.

\subsection {Newtonian polytropes}\label{sec:newtpoly}
In Newtonian physics,
the mass-radius relationship is given as a function of the polytropic
index $n$ by
\begin{equation}
R_{\rm NS}\propto M_{\rm NS}^{(1-n)/(3-n)}.
\end{equation}
For a given value of the parameter $\kappa$ in
$P=\kappa\rho_0^{(1+1/n)}$, the familiar result is that NSs with $n=1$ have a
uniform radius independent of their mass.  NSs with $n<1$ have radii
which decrease as the mass decreases, whereas those with $n>1$ increase
in size as they lose mass.  We see that if the NS loses mass at a rate 
$\dot{m}\equiv -\dot{M}_{\rm NS}$ (implying $\dot{m}>0$ for the case
of interest), the change in the radius is given by
\begin{equation}
\frac{\dot{R}_{\rm NS}}{R_{\rm
    NS}}=\frac{n-1}{3-n}\frac{\dot{m}}{M_{\rm NS}}.\label{eq:rpoly} 
\end{equation}

Once mass transfer begins, the NS will shed mass, and in doing so,
lose both energy and angular momentum. This will affect the binary
orbit, in a way which depends on both the magnitude and fate of the 
angular momentum of the ejected material.

For conservative mass transfer, 
the orbital angular momentum $J$ and the total binary mass $M_T\equiv
M_{\rm NS}+M_{\rm BH}=(1+q)M_{\rm NS}/q$ are conserved
globally, as mass is transferred from the NS to the BH.  We assume
that the orbit remains circular.  Since $q=M_{\rm NS}/(M_T-M_{\rm
  NS})$, we see that
\begin{equation}
\frac{\dot{q}}{q}=-(1+q)\frac{\dot{m}}{M_{\rm NS}}.
\end{equation}
From the angular momentum of a circular orbit,
\begin{equation}
J=M_{\rm NS}M_{\rm BH}\sqrt{\frac{Ga}{M_T}},\label{eq:jtot}
\end{equation}
we find that
\begin{eqnarray}
a&=&\frac{M_TJ^2}{G}\left[M_{\rm NS}(M_T-M_{\rm
    NS})\right]^{-2}\nonumber\\&=&\frac{J^2}{G}M_{\rm NS}^{-3}q(1+q),\\
\dot{a}/a&=&2(1-q)\frac{\dot{m}}{M_{\rm NS}}.\label{eq:aadot}
\end{eqnarray}
The Roche-lobe radius $R_r$ changes during the process as well.  Taking the
logarithmic time derivative of Eq.~(\ref{eq:rroche}) and combining
with Eq.~(\ref{eq:aadot}), we find
\begin{eqnarray}
\frac{\dot{R}_{r}}{R_r}&=&\frac{\dot{a}}{a}+\frac{\dot{q}}{3q(1+q)}\nonumber\\
&=&\frac{\dot{a}}{a}-\frac{\dot{m}}{3M_{\rm
    NS}}=\frac{5-6q}{3}\frac{\dot{m}}{M_{\rm NS}}. \label{eq:aadotrrdot}
\end{eqnarray}
Combining this final expression with Eq.~(\ref{eq:rpoly}), we see
that mass transfer will be unstable if 
$\dot{R}_{\rm NS}/R_{\rm NS} > \dot{R}_r/R_r$, or equivalently,
\begin{eqnarray}
n &>& \frac{9-9q}{4-3q},\label{eq:nqnewt}\\
q &>& \frac{9-4n}{9-3n},
\end{eqnarray}
or in terms of the adiabatic index $\Gamma\equiv 1+1/n$, mass transfer
is unstable if
\begin{eqnarray}
\Gamma &<& \frac{13-12q}{9-9q},\label{eq:gamqnewt}\\
q &>& \frac{9\Gamma-13}{9\Gamma-12}.
\end{eqnarray}
In particular, the critical mass ratio for unstable mass transfer for
some polytropic indices commonly used in numerical calculations are
\begin{eqnarray}
n=1/2~(\Gamma=3) &:& q>14/15,\\
n=1~(\Gamma=2) &:& q>5/6,\\
n=3/2~(\Gamma=5/3) &:& q>2/3.
\end{eqnarray}
As a general rule, for polytropic indices $n<3/2$, mass transfer is
stable only for systems with components of similar mass. For $n>3/2$,
the critical mass ratio drops quickly, down to the limiting case
$n=9/4$, the softest polytropic EOS for which stable mass transfer can
ever occur, at which point $q=0$.

\subsection{Relativistic polytropes}\label{sec:relpoly}
Relativistic polytropes are not self-similar for a fixed value of the
polytropic index; as the mass decreases, non-linear gravitational
effects become weaker, and the star's scale-free density profile grows
in size.  It is straightforward to incorporate this into our
discussion of mass transfer.  We recast the mass-radius relationship
in the form
\begin{equation}
R_{\rm NS}\equiv \xi M_{\rm NS}^{(1-n)/(3-n)}f({\cal C}),\label{eq:massrad}
\end{equation}
where $\xi$ sets the physical scale of the mass-radius relation,
${\cal C}\equiv M/R$ is the compactness of a spherical star, and  
$f({\cal C})$ accounts for the relativistic corrections to the
mass-radius relation.   
Since relativistic corrections effectively increase the strength of
gravity, $f({\cal C})$ must be a monotonically decreasing function of
compactness. As ${\cal C}\rightarrow 0$, $\xi$ approaches 
the proper Newtonian value, $\xi=\xi_N$, and  $f(0)\rightarrow1$. 
Taking a logarithmic derivative of Eq.~(\ref{eq:massrad}) shows us that
\begin{equation}
\frac{\dot{R}_{\rm NS}}{R_{\rm NS}}=
\frac{\dot{f}}{f}+\frac{n-1}{3-n}\frac{\dot{m}}{M_{\rm NS}}.
\end{equation}
But we know
\begin{eqnarray}
\frac{\dot{f}}{f}=\frac{1}{f}\frac{\partial f}{\partial {\cal
    C}}\dot{{\cal C}}&=& 
\frac{1}{f}\frac{\partial f}{\partial {\cal C}}\left[\frac{\dot{M}_{\rm
      NS}}{R_{\rm NS}}-\frac{M\dot{R}_{\rm
      NS}}{R_{\rm NS}^2}\right]\nonumber\\&=&
-\frac{{\cal C}}{f}\frac{\partial f}{\partial {\cal
    C}}\left[\frac{\dot{m}}{M_{\rm 
	NS}}+ \frac{\dot{R}_{\rm NS}}{R_{\rm NS}}\right],
\end{eqnarray}
so we conclude that
\begin{equation}
\frac{\dot{R}_{\rm NS}}{R_{\rm NS}}=
\left[\frac{n-1}{3-n}-\frac{2}{3-n}\frac{{\cal C}}{f}\frac{\partial
    f}{\partial {\cal C}}\left(1+\frac{{\cal C}}{f}\frac{\partial
    f}{\partial {\cal C}}\right)^{-1}\right]\frac{\dot{m}}{M_{\rm NS}}.
\label{eq:rdotnsrns}
\end{equation}
From Eq.~(\ref{eq:nqnewt}), we 
see that the critical mass ratio for instability becomes
\begin{equation}
q_c=\frac{9-4n}{9-3n}+\frac{1}{3-n}\frac{{\cal C}}{f}\frac{\partial f}{\partial
  {\cal C}}\left(1+\frac{{\cal C}}{f}\frac{\partial
    f}{\partial {\cal C}}\right)^{-1},
\end{equation}
where the second term is always negative, indicating that more
compact configurations are more unstable against mass transfer.

\begin{figure}[ht]
\centering \leavevmode \epsfxsize=\columnwidth \epsfbox{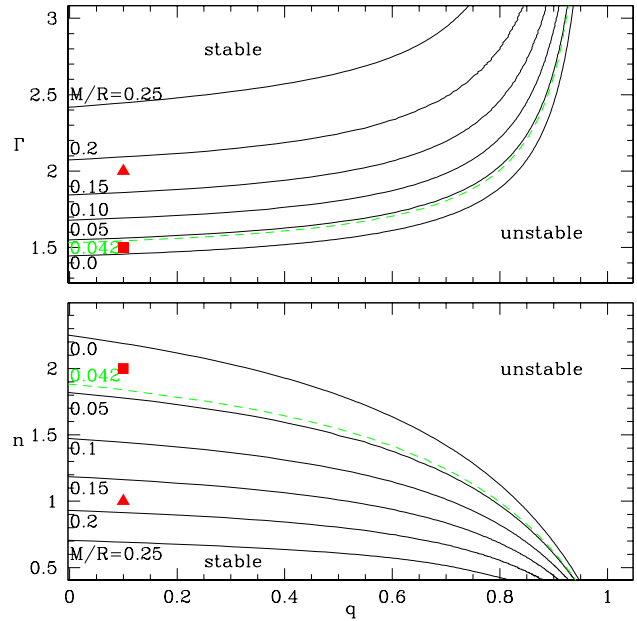}
\caption{Critical compactness for determining the stability of 
mass transfer from a relativistic polytropic NS companion as a
function of the binary mass ratio, 
defined in terms of the polytropic index $n$
(bottom), or the adiabatic index $\Gamma\equiv 1+1/n$.  Mass transfer
is unstable to the right of the curve, stable to the left. The square
and triangle represent the positions of the two models we are evolving
dynamically.} 
\label{fig:ngamqc}
\end{figure}

In Figure~\ref{fig:ngamqc}, we show the critical compactness values
separating stable and unstable mass transfer for relativistic
polytropes, as a function of the binary mass ratio and the polytropic
(bottom panels) and adiabatic (top panels) indices.  As a general
rule, as the compactness of the NS increases, mass transfer is more
likely to be unstable. The Newtonian curves (${\cal C}=0$) have an analytic
form given by Eqs.~(\protect\ref{eq:nqnewt}) and (\protect\ref{eq:gamqnewt}).
 Also shown are the two models we evolve
dynamically in this paper, both with ${\cal C}=0.042$.  The case with
$n=1 (\Gamma=2)$, 
shown as a triangle, would be expected to demonstrate stable mass
transfer in the highly viscous regime, but as we show in
Sec.~\ref{sec:gam2}, the true situation is nowhere near this simple
for typical NSs.
The case with $n=2 (\Gamma=1.5)$, shown as a square, is expected to
lead to unstable mass transfer.

\begin{figure}[ht]
\centering \leavevmode \epsfxsize=\columnwidth \epsfbox{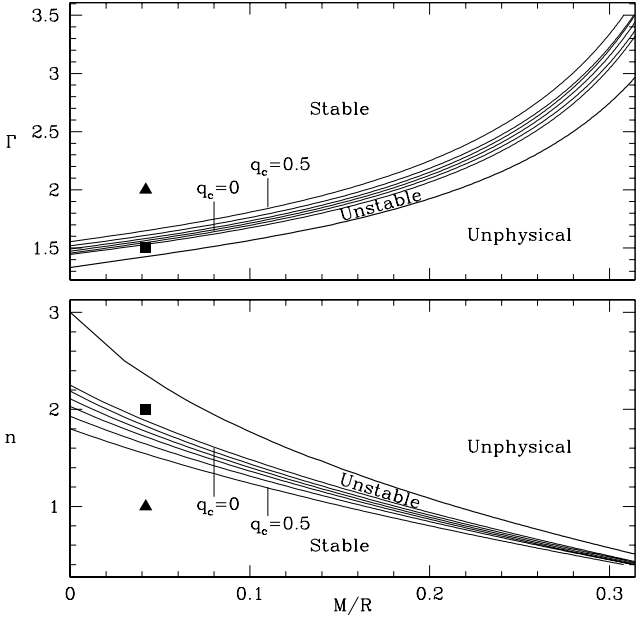}
\caption{Critical mass ratio for determining the stability of 
mass transfer from a relativistic polytropic NS companion as a
function of the NS compactness, 
defined in terms of the polytropic index $n$
(bottom), or the adiabatic index $\Gamma\equiv 1+1/n$.  Mass transfer
is unstable to the right of the curve, stable to the left. The heavy
solid lines depict the maximum possible compaction for a given
polytropic EOS.  The square
and triangle represent the positions of the two models we are evolving
dynamically.} 
\label{fig:ngamqc2}
\end{figure}

Representing the same results against compactness, rather than mass
ratio, shows how little parameter space is available for unstable mass
transfer in the viscous limit. In Figure~\ref{fig:ngamqc2}, we show
the critical binary 
mass ratio for stable or unstable mass transfer as a function of the NS
compactness and polytropic (bottom) or adiabatic (top) index.  We see
that for a given compactness, there is a rather limited range of
polytropic indices which can produce unstable mass transfer.  The
heavy lines show the maximum possible compactness allowed for a given
polytropic EOS; no model to the right of those curves can be
constructed.  The models that we evolve dynamically are shown as well.
As noted in Sec.~\ref{sec:gam2}, many models expected to undergo stable mass
transfer in the viscous limit are extremely unstable during mass
transfer when viscosity is not a dominant driver of the evolution
\cite{BC,Koch}, so the true parameter space for instability is actually
much larger for typical NSs than analytic models would otherwise predict.

\subsection{Time evolution}
We describe here a crude treatment of the dynamics of
mass transfer in the presence of gravitational radiation-reaction
losses, using a number of simplifying assumptions. Most important
among these will be treating the problem in the quasi-equilibrium regime. 
 
We will study only conservative mass transfer, in the highly viscous limit.
The evolution of the
binary separation can be derived from the expression for
the total angular momentum for a circular orbit, Eq.~(\ref{eq:jtot}).
Making no assumptions about
the evolution of the system angular momentum, we find
\begin{equation}\label{eq:jdotcons}
\frac{\dot{J}}{J}=\frac{\dot{M_{\rm NS}}}{M_{\rm
    NS}}+\frac{\dot{M_{\rm BH}}}{M_{\rm BH}}+\frac{\dot{a}}{2a} 
=(q-1)\frac{\dot{m}}{M_{\rm NS}}+\frac{\dot{a}}{2a}.
\end{equation}
Setting the RHS equal to zero gives the standard expression for
conservative mass transfer, Eq.~(\ref{eq:aadot}).  
For the case of a relativistic binary, we
know angular momentum will not be conserved. Instead, angular momentum
is radiated away from the system in gravitational waves.  As we are
assuming circular orbits, we will make use of the standard angular
momentum loss rate,
\begin{equation}\label{eq:jdotrr}
\frac{\dot{J}}{J}=-\frac{32}{5}\frac{M_{\rm NS}M_{\rm BH}M_T}{a^4}.
\end{equation}
From our two angular momentum relations, we can derive a relation
linking the mass loss rate to the change in binary separation, and
determine the time-averaged mass loss rate.  To do so, 
we will make a very simple
assumption: that mass loss takes place at the location where we
predict Roche-lobe overflow to occur.
 
Solving Eq.~(\ref{eq:aadotrrdot}) for the binary separation and assuming
$R_r=R_{\rm NS}$, we can insert Eq.~(\ref{eq:rdotnsrns}) and find
\begin{eqnarray}
\frac{\dot{a}}{a}\!\!&=&\frac{2n}{9-3n}\frac{\dot{m}}{M_{\rm
    NS}}+\frac{\dot{f}}{f}\nonumber\\
&=&\!\!\!\!\frac{2}{3-n}\left[\frac{n}{3}-\frac{{\cal C}}{f}\frac{\partial
    f}{\partial {\cal C}}\left(1+\frac{{\cal C}}{f}\frac{\partial
    f}{\partial {\cal C}}\right)^{-1}\right]\frac{\dot{m}}{M_{\rm NS}}.
\end{eqnarray}
Plugging this result into Eq.~(\ref{eq:jdotcons}), we find
\begin{equation}
\frac{\dot{J}}{J}=-\left[\frac{9-4n}{9-3n}-q+
\frac{1}{3-n}\frac{{\cal C}}{f}\frac{\partial
    f}{\partial {\cal C}}\left(1+\frac{{\cal C}}{f}\frac{\partial
    f}{\partial {\cal C}}\right)^{-1}\right]\frac{\dot{m}}{M_{\rm NS}}.
\end{equation}
Re-expressing the radiation-reaction angular momentum loss rate,
Eq.~(\ref{eq:jdotrr}), for a binary located at the Roche-lobe overflow
point, we find
\begin{eqnarray}
\frac{\dot{J}}{J}\!\!&=&-\frac{32}{5}\frac{M_{\rm NS}M_{\rm
  BH}M_T}{\left(0.46 R_{\rm NS} 
  M^{1/3}m_1^{-1/3}\right)^4}\nonumber\\&=&\!\!\!-143 \xi^{-4}\!f({\cal C})^{-4}
(\!M_T\!-M_{\rm NS})\!M_T^{-1/3}\!M_{\rm NS}^\frac{9+5n}{9-3n}.
\end{eqnarray}
We are now in a position to analyze the mass-transfer process from its
onset until the final fate of the binary.
To do so, we will make a few simplifying assumptions.  First, we
assume the NS is much less massive than the BH, such that $M_{\rm BH}\sim
M_T$ and $q\ll 1$.  Also, we will ignore the relativistic changes to
the Newtonian 
power-law mass-radius relation, which has the effect of setting $f=1$
uniformly (and thus $\partial f/\partial {\cal C}=0$).  In general,
relativistic polytropes expand more 
than their Newtonian analogues during mass loss, which increases the
amount of orbital expansion seen for a given amount of mass loss.
This statement would imply that a given amount of mass loss leads to a
greater loss of angular momentum, or conversely, that for a fixed
angular momentum loss rate we would see a slight suppression of the
mass loss rate.
Under these assumptions, we find
\begin{eqnarray}
\frac{\dot{m}}{M_{\rm NS}}&=&143 \xi^{-4}M_T^{2/3}\frac{9-3n}{9-4n}
M_{\rm NS}^\frac{9+5n}{9-3n}\nonumber\\&=&
\frac{9-3n}{9-4n}\left(\frac{M_{\rm NS}}{M_{\rm
    NS}(0)}\right)^{\frac{9+5n}{9-3n}}\frac{1}{2t_{GW0}},\label{eq:dotmmns} 
\end{eqnarray}
where we define $t_{GW0}$ as the infall timescale
when the NS first reaches the onset of mass transfer, as defined by
Eq.~(\ref{eq:tgw}), and where $M_{\rm NS}(0)$ is the NS mass at this time.
This equation has the solution
\begin{equation}
\frac{M_{\rm NS}(t)}{M_{\rm
    NS}(0)}=\left(1.0+\frac{9+5n}{9-4n}\frac{t}{2t_{\rm
    GW0}}\right)^{-\frac{9-3n}{9+5n}},
\end{equation}
with asymptotic behavior
\begin{eqnarray}
M_{\rm NS}(t)&\propto& t^{-\frac{9-3n}{9+5n}},\\
\dot{m}&\propto& t^{-\frac{18+2n}{9+5n}}.
\end{eqnarray}
Considering some familiar polytropic EOS, we find:
\begin{eqnarray}
n=1.5~(\Gamma\!=5/3)&:&\!\!M\propto t^{-3/11};\dot{m}\propto
t^{-14/11},\label{eq:spz}\\
n=1.0~(\Gamma=2)&:&\!\!M\propto t^{-3/7};\dot{m}\propto t^{-10/7},\\
n=0.5~(\Gamma=3)&:&\!\!M\propto t^{-15/23};\dot{m}\propto t^{-38/23},\\
n=0~(\Gamma\rightarrow\infty)&:&\!\!M\propto t^{-1};\dot{m}\propto t^{-2}.
\end{eqnarray}
Eq.~(\ref{eq:spz}) recovers the special value found by \cite{PZ}, in the limit
$q\ll 1$.  We note that the typical rate found here satisfies
$\dot{M}_{\rm NS}\sim M_{\rm NS}/t_{GW0}$ [cf.,
  Eq.~(\ref{eq:dotmmns})], while our dynamical simulations for cases in
which viscosity is not important show $\dot{M}_{\rm NS}\gg M_{\rm
  NS}/t_{GW0}$. 

\section{Symmetries}\label{app:symm}
When constructing quasi-equilibrium configurations in either Newtonian
or relativistic gravity, it is important to take note of the various
symmetries present in the relevant equations.  These symmetries can
either be enforced numerically, to save computational resources, or be
used as a check to make sure that a numerical code is producing
physically valid results.

For quasi-equilibrium binary BHNS systems, virtually all
gravitational formalisms will produce a configuration that is
equatorially symmetric so long as the NS spin axes are parallel to the
orbital angular momentum axis, regardless of whether the NS is
corotating, counterrotating, or 
irrotational.   If we fix the z-axis to be parallel to the various angular
momenta mentioned above,
we find that the transformation $z\rightarrow -z$, with a
corresponding reflection for all vector and tensor quantities, leaves
the hydrostatic equations and the metric invariant, and is compatible
with the velocity field of the initial configuration as well.
For quasi-equilibrium configurations evaluated using full GR 
in the Kerr-Schild metric, this is the only symmetry plane of note present
in the initial data. In our calculations, this symmetry is enforced at
the code level: each SPH particle is treated as if it were a pair of
particles, each of half the total mass, with one copy lying above the
equatorial plane and one below.  All spectral decompositions include
this symmetry as well, setting to zero all spherical harmonics that
are incompatible with a fully symmetric description.

We show here that there is an additional
symmetry plane for the case of equilibrium BHNS binaries in
conformally flat gravity, because the
formalism is time-symmetric.  Directions are defined such that the
axis of separation between the BH and NS lies along the x-axis, and
the orbital angular momentum points in the z-direction.  Thus, the
y-direction represents the direction of motion for each object.
Reversing the time direction requires us to perform two operations in
order to maintain invariance.  First, we must invert all vector
quantities, most notably the velocity and the shift vector.  Second,
we must perform an inversion in the {\em y-direction}, to account for
the inversion of our angular coordinate $\phi\rightarrow -\phi$.  We
find the following relations are compatible with the initial velocity
field, and leave our hydrostatic equations invariant:
\begin{eqnarray}
{\rm Scalars}&:& f(x,y,z,t)=f(x,-y,z,-t),\\
{\rm Vectors}&:& [v^x,v^y,v^z](x,y,z,t)=\nonumber\\&&~~~~-[v^x,-v^y,v^z](x,-y,z,-t).
\end{eqnarray}
Evaluating the expressions above at $t=0$ yields the symmetries in the
y-direction for our initial data.  This symmetry can be extended
to tensor quantities as well: two index tensor elements satisfy the relation
$T_{ij}(x,-y,z)=(-1)^{1+N_y}T_{ij}(x,y,z)$, where $N_y$ is the
number of ``y'' indices present for a given element.
The additional factor of $-1$ is necessary
to describe the inversion properties under time symmetry.  Equatorial
symmetry can be described in the same language, 
$T_{ijk...}(x,y,-z)=(-1)^{N_z}T_{ijk...}(x,y,z)$. 

It is straightforward to check that these symmetries are compatible
with all equations governing the construction of quasi-equilibrium NS
binaries in CF gravity.  To do so, we note that any tensorial
operation on fields satisfying these symmetries will maintain these
symmetries, including gradients and inner products.

First, we note that the matter sources in the equations for the
conformal factor and lapse function, Eqs.~(\ref{eq:psi}) and
(\ref{eq:alphapsi}), 
$E$, $P$, and $S$, as scalars, are symmetric in
both $y$ and $z$.  
The same pattern follows for scalar densities and quantities like $u^0$.
As an immediate consequence, the lapse and conformal factor share the
same symmetries, since the only other source terms involve $K_{ij}K^{ij}$,
itself a scalar.

The equilibrium 
velocity field is $\vec{v}=\vec{\Omega}\times\vec{r}=\allowbreak [-\Omega y,
  \Omega x, 0]$, which satisfies
\begin{equation}
[v^x,v^y,v^z](x,-y,z)=-[v^x,-v^y,v^z](x,y,z).
\end{equation}
Equatorial symmetry is satisfied trivially.
In the CF case, the BH component of the shift vector is zero, and also
satisfies the symmetry relations trivially.  
Since the shift equation, Eq.~(\ref{eq:shift}) depends only on the
velocity field and other tensors, we can extend the symmetry to
describe all quantities present in the equation.
This statement cannot be
made for the Kerr-Schild metric, and serves as the simplest
example of how time-symmetry fails.  For the Kerr-Schild case,
\begin{eqnarray}
\beta^i_{\rm BH-KS}&\propto& x^i\rightarrow\nonumber\\
&&\!\!\!\!\!\!\!\!\!\!\!\!\!\!\!\!\!\!\!\!\!\!\!\!
[\beta^x,\beta^y,\beta^z](x,-y,z)=
[\beta^x,-\beta^y,\beta^z](x,y,z).
\end{eqnarray}
Thus, the BH contribution to the shift satisfies a different
symmetry relation than the initial velocity field, and there is no
global symmetry present.  The physical reasons underlying this fact
are discussed in more detail in Appendix~E of \cite{YCSB}.

\section{Preserving Stationary Equilibrium During CF Evolution}
\label{appendix:invariance}

We show here that the CF evolution equations maintain strict
stationary equilibrium for initial data satisfying 
both the thin-sandwich equations and the integrated Euler equation.
That is, if we have an initially uniformly rotating 
matter configuration that satisfies
the condition $h/u^0 = C$, where $C$ is constant throughout space
at $t=0$, and if the gravitational field satisfies the CTS equations,
Eqs.~(\ref{eq:shift}), (\ref{eq:psi}), and (\ref{eq:alphapsi}),
the time derivatives of all quantities go to zero.  Note that under
these assumptions, this statement applies to both Eulerian and
Lagrangian time derivatives, related by the expression $d/dt=\partial/\partial
t+v^j\partial/\partial x^j$ since the difference term goes to zero when
$v^j=0$, which applies in the corotating frame.

First, we note that the continuity equation, Eq.~(\ref{eq:cont}), and
energy equation, Eq.~(\ref{eq:destardt}), are trivially conserved when
$v^i=0$, implying that $\rho_*$ and $e_*$ are conserved automatically.

The only evolution equation that requires a more thorough look is the
Euler equation, Eq.~(\ref{eq:dutildedt}), which we re-express for
convenience in the equivalent form
\begin{equation}
\frac{d\tilde{u}_i}{dt}=-\frac{\partial_i P}{\rho_* u^0}-
\alpha hu^0\partial_i \alpha+\tilde{u}_j\partial_i
\beta^j+\frac{2\tilde{u}_k\tilde{u}_k}{hu^0 \psi^5}\partial_i \psi.
\end{equation}

We can show that the RHS of this expression is zero by starting from
the integrated Euler equation, which implies that
\begin{equation}
\frac{\partial_i h}{u^0}-\frac{h\partial_i u^0}{(u^0)^2}=0,
\end{equation}
From the relativistic Gibbs-Duhem relation, we know that \cite{GGTMB}
\begin{equation}
\frac{\partial_i h}{u^0}=\frac{\partial_i P}{\rho_0 u^0}.
\end{equation}
 The gradient of $u^0$ is determined from the normalization
 $u_au^a=u_0u^0=-1$, where we make use of the fact that $u^i=0$.
From this, we conclude
\begin{equation}
u^0=(\alpha^2-\psi^4\delta_{ij}\beta^i\beta^j)^{-1/2},
\end{equation}
and differentiating,
\begin{equation}\label{eq:du0dt}
\partial_iu^0=-(u^0)^3(\alpha\partial_i\alpha
-\psi^4\delta_{jk}\beta^j\partial_i\beta^k
-2\psi^3\delta_{jk}\beta^j\beta^k\partial_i\psi).
\end{equation}
Inserting the expression $u_i=g_{0i}u^0=-\psi^4u^0\delta_{ij}\beta^j$ for a
configuration with $u^i=0$, we see that
\begin{equation}
\frac{h\partial_i u^0}{(u^0)^2}=-\left(\alpha hu^0\partial_i\alpha
+\tilde{u}_k\partial_i\beta^k
-\frac{2\tilde{u}_k\tilde{u}_k}{hu^0 \psi^5}\partial_i\psi\right),
\end{equation}
and combining terms,
\begin{eqnarray}
0&=&\partial_i\left(\frac{h}{u^0}\right)=\frac{\partial_i
  h}{u^0}-\frac{h\partial_i u^0}{(u^0)^2}\nonumber\\
&=&\frac{\partial_i P}{\rho_0
  u^0}+ 
\left(\alpha hu^0\partial_i\alpha
+\tilde{u}_k\partial_i\beta^k
-\frac{2\tilde{u}_k\tilde{u}_k}{hu^0 \psi^5}\partial_i\psi\right)\nonumber\\
&=&-\frac{d\tilde{u}_i}{dt}=0.
\end{eqnarray}
Thus, the RHS of the Euler equation, Eq.~(\ref{eq:dutildedt}), 
is zero under these assumptions, and $\tilde{u}_i$ is also invariant.

We now consider the field equations for the CTS initial data.
The fields are described by a set of linked elliptic
equations, Eqs.~(\ref{eq:shift}), (\ref{eq:psi}), and (\ref{eq:alphapsi}), 
whose source terms involve the fields themselves, as well
as three quantities:
\begin{eqnarray}
E&=&(\alpha u^0)^2[\frac{\Gamma P}{\Gamma-1}+\rho_0]-P,\\
U^i&=&\frac{\delta^{ij}u_j}{\psi^4\gamma_n (1+\Gamma\epsilon)},\\
S&=&3P+\psi^4(E+P)\delta_{ij}U^iU^j,
\end{eqnarray}
where $\rho_0$, $\epsilon$, and $P$ are the standard relativistic mass
density, internal energy density, and pressure, respectively.  The
Lorentz factor $\gamma_n\equiv \alpha u^0$, can be solved implicitly,
from Eq.~(\ref{eq:gamman}).
Coupled with our field equations, we have a set of six completely
linked equations for six variables ($\gamma_n$, $\alpha$, $\psi$,
$\beta^i$) and our conserved matter quantities ($\rho_*$, $e_*$,
$\tilde{u}_i$).  So long as a unique solution exists for our choice of
matter variables, we know that this solution will remain invariant so
long as we choose our elliptic equation boundary conditions to be
invariant.
Thus, the RHS of all our evolution equations will remain zero, and the
matter configuration will remain in equilibrium.

\end{document}